\documentclass[nofootinbib,twocolumn,showpacs,,eqsecnum,preprintnumbers,amsmath,amssymb]{revtex4}
\usepackage{graphicx}
\usepackage{ulem,color,graphics}
\usepackage{dcolumn}
\usepackage{bm}
\usepackage{wrapfig}
\usepackage{sidecap}
\begin{document}
\include{def}
\def\etal{{\it et al.}}
\def\go{\rightarrow  }
\def\be{\begin{equation}}
\def\ee{\end{equation}}
\def\br{\begin{eqnarray}}
\def\er{\end{eqnarray}}
\def\brn{\begin{eqnarray*}}
\def\ern{\end{eqnarray*}}
\def\rf#1{{(\ref{#1})}}
\def\a {{\alpha}}
\def\b {{\beta}}
\def\e {{\epsilon}}
\def\k {{\kappa}}
\def\s {{\sigma}}
\def\w {{\omega}}
\def\sss{\scriptscriptstyle}
\def\nn{\nonumber}
\def\ie{{\it i.e., }}
\def\x{\times}
\def\F {{{\cal F}}}
\def\O {{{\cal O}}}
\def\M {{{\cal M}}}
\def\T {{{\cal T}}}
\def\L {{{\cal L}}}
\def\pb {{\bf p}}
\def\qb {{\bf q}}
\def\rb {{\bf r}}
\def\kb {{\bf k}}
\def\ket#1{|#1 \rangle}
\def\bra#1{\langle #1|}
\def\Ket#1{||#1 \rangle}
\def\Bra#1{\langle #1||}
\def\up{u_{{\rm p}}}
\def\vp{v_{{\rm p}}}
\def\un{u_{{\rm n}}}
\def\vn{v_{{\rm n}}}
\def\etc{ {\it etc}}
\def\E {{{\cal E}}}
\def\ga{\overline{g}_{\mbox{\tiny A}}}
\def\gv{\overline{g}_{\mbox{\tiny V}}}
\def\gp{\overline{g}_{\mbox{\tiny P}}}
\def\gpa{\overline{g}_{\mbox{\tiny P1}}}
\def\gpb{\overline{g}_{\mbox{\tiny P2}}}
\def\gw{\overline{g}_{\mbox{\tiny W}}}
\def\gm{\overline{g}_{\mbox{\tiny M}}}
\def\gA{g_{\mbox{\tiny A}}}
\def\rA{r_{\mbox{\tiny A}}}
\def\gAeff{g_{\mbox{\tiny A}}^{\mbox{\tiny eff}}}
\def\gV{g_{\mbox{\tiny V}}}
\def\gM{g_{\mbox{\tiny M}}}
\def\gS{g_{\mbox{\tiny S}}}
\def\gT{g_{\mbox{\tiny T}}}
\def\gP{g_{\mbox{\tiny P}}}
\def\gPeff{g_{\mbox{\tiny P,eff}}}
\def\GA{G_{\mbox{\tiny A}}}
\def\GV{G_{\mbox{\tiny V}}}
\def\GF{G_{\mbox{\tiny F}}}
\def\gvs{g_{\sss{V}}}
\def\gas{g_{\sss{A}}}
\newcommand{\Mass}{\mathrm{M}}
\newcommand{\mass}{\mathrm{m}}
\def\Ss{\mathscr{S}}
\def\Ys{\mathscr{Y}}
\def\Os{\mathscr{O}}
\def\sq{{\sqrt 2}}
\def\lb {{\bf l}}
\def\Jb {{\bf J}}
\def\mbs{\mbox{\boldmath$\sigma$}}
\def\mbn{\mbox{\boldmath$\nabla$}}
\def\Ob {{\bf O}}
\def\kr {{\bf k}\cdot{\bf r}}
\def\qr {{\bf q}\cdot{\bf r}}
\def\threej#1#2#3#4#5#6{\left(\negthinspace\begin{array}{ccc}
#1&#2&#3\\#4&#5&#6\end{array}\right)}
\def\sixj#1#2#3#4#5#6{\left\{\negthinspace\begin{array}{ccc}
#1&#2&#3\\#4&#5&#6\end{array}\right\}}
\def\bit{\begin{itemize}}
\def\eit{\end{itemize}}
\def\lsim{\:\raisebox{-0.5ex}{$\stackrel{\textstyle<}{\sim}$}\:}
\def\gsim{\:\raisebox{-0.5ex}{$\stackrel{\textstyle>}{\sim}$}\:}
\newcommand{\red}[1]{\textcolor[rgb]{1.00,0.00,0.00}{\sf#1}}

\preprint{}

\title{Neutrino and antineutrino charge-exchange reactions on $^{12}$C}

\author{A. R. Samana$^{1,2}$, F. Krmpoti\'c$^{3}$, N. Paar$^{4}$, and C. A. Bertulani$^{1}$}
\affiliation{$^1$ Department of Physics, Texas A\&M University Commerce,
P.O.3011 Commerce, 75429 TX, USA}%
\affiliation{$^2$
Departamento de Ci\^encias Exactas e Tecnol\'ogicas,
Universidade Estadual  de Santa Cruz,
CEP 45662-000 Ilheús, Bahia-BA, Brazil}%
\affiliation{$^3$Instituto de F\'isica La Plata, CONICET,
 Facultad de Ciencias Astron\'omicas y Geof\'isicas,\\
 Universidad Nacional de La Plata, 1900 La Plata, Argentina,}%
\affiliation{$^4$Physics department, Faculty of Science, University of Zagreb, Croatia}

\date{\today}
\begin{abstract}
 We extend the  formalism of weak interaction processes, obtaining
 new  expressions
for the transition rates,  which  greatly facilitate numerical calculations,
 both for neutrino-nucleus reactions and muon capture. Explicit violation of CVC
hypothesis by the Coulomb field, as well as development of a sum rule approach
for the inclusive cross sections have been worked out. We have done
a thorough  study of exclusive (ground state)
 properties of  $^{12}$B and
$^{12}$N  within  the projected quasiparticle random phase
approximation (PQRPA). Good agreement with experimental data achieved  in this way put in evidence the limitations
of standard  RPA and the QRPA models, which come from the inability of
the RPA in opening the $p_{3/2}$ shell, and from the
non-conservation of the number of particles in the QRPA.
The inclusive neutrino/antineutrino ($\nu/\tilde{\nu}$) reactions
  $^{12}$C($\nu,e^-)^{12}$N and $^{12}$C($\tilde{\nu},e^+)^{12}$B
are calculated within both the PQRPA,  and the relativistic QRPA (RQRPA).
It is found that the magnitudes of the resulting cross-sections:
i) are  close to the sum-rule limit
at low energy, but significantly smaller than this limit
at high energies both for $\nu$ and $\tilde{\nu}$,
ii) they steadily increase when the  size of the configuration space is augmented,
and particulary for $\nu/\tilde{\nu}$ energies $> 200$ MeV,
and iii) converge  for sufficiently large configuration space and
  final state spin. The quasi-elastic   $^{12}$C($\nu,\mu^-)^{12}$N
   cross section   recently measured
   in the MiniBooNE experiment is briefly discussed.
We study the decomposition of the inclusive
cross-section based on the degree of forbiddenness  of different
multipoles. A few words are dedicated
to the $\nu/\tilde{\nu}$-$^{12}$C charge-exchange
reactions related with astrophysical applications.
\end{abstract}
\pacs{23.40.-s, 25.30.Pt, 26.50.+x}
\maketitle
\section{Introduction}\label{Sec1}
 The massiveness
of neutrinos and the related oscillations are strongly sustained by
many experimental works involving atmospheric, solar, reactor and
accelerator neutrinos
~\cite{Ath96,Ath98,Agu01,Fuk98,Aha05,Ara04,Ahn03}.
 The  subsequent experimental goal   is to determine precisely the
various parameters of the
Pontecorvo-Maki-Nakagawa-Sakata (PMNS) neutrino mass matrix, absolute
masses of different flavors of neutrinos, CP violation in neutrino
sector, etc. To address  these problems several
 analyses of neutrino oscillation data are
presently going on. At the same time, several
experiments  presently collect data, and others are planned.
Accelerator experiments, experiments with neutrinos from $\nu$-factories, $\beta$-beams, etc.,
are also planned and designed, as well as some experiments with
natural $\nu$-sources like solar neutrinos, atmospheric neutrinos,
or antineutrinos from nuclear reactors.

The neutrino-nucleus scattering on $^{12}$C
is important because this nucleus is a component of many
liquid scintillator detectors. Experiments such as LSND~\cite{Ath96,Ath98},
 KARMEN~\cite{Mas98,Arm02}, and
LAMPF~\cite{All90,Kra92}  have used
$^{12}$C to search for neutrino oscillations, and for measuring
neutrino-nucleus cross sections.
Present atmospheric and accelerator based
neutrino oscillation experiments  also involve $^{12}$C, and operate at
neutrino energies $E_\nu\sim 1$ GeV in order to access the relevant
regions of oscillation parameter space.
This is the case of the SciBar detector~\cite{SciBar}, where the molecule $C_8H_8$
is involved, and the  MiniBooNE detector ~\cite{MiniBooNE}, which uses the light mineral oil
containing the molecule $CH_2$.
The $^{12}$C target will be used in
several planned experiments, such
as the spallation neutron source (SNS) at Oak Ridge National
Laboratory (ORNL)~\cite{Efr05}, and the LVD (Large Volume Detector)
experiment~\cite{Aga07}, developed by the INFN in Gran Sasso.

For the planned experimental searches  of  supernovae neutrino signals, which
involve $^{12}$C as scintillator liquid detector,  the precise
knowledge of neutrino cross sections  of $^{12}$N and $^{12}$B ground-states,
 \ie of $\sigma_{e^-}(E_\nu,1^+_{1})$, and
$\sigma_{e^+}(E_{\tilde{\nu}},1^+_{1})$  is very important. In fact,
in the LVD experiment~\cite{Aga07} the number of events detected
during the supernova explosion are estimated  by
convoluting  the neutrino supernova flux with: i)  the
interaction cross sections, ii) the efficiency of the detector, and
iii) the number of target nuclei. For the carbon content of
the LVD detector have been used so far
$\sigma_{e^-}(E_\nu,1^+_{1})$, and
$\sigma_{e^+}(E_{\tilde{\nu}},1^+_{1})$, as obtained from the Elementary Particle
Treatment (EPT)~\cite{Fuk88}.
Moreover, as an update of the LVD experiment related to supernovae neutrinos
detection (where $^{12}$C will also be employed), there is ongoing
design study  concerning large size scintillator detectors, called
LAGUNA, where a 50 kt scintillator LENA is being considered
\cite{Aut07}.

On the other hand,  as the $^{12}$C nucleus forms one  of the
onion-like shells of a large star before collapse, it is also
important for astrophysics studies. Concomitantly, several
authors~\cite{Lun03,Dig03,Aga07,Dua07,Das08,Das10,Mez10}
 have recently stressed
the importance of  measuring supernova neutrino oscillations.
They claim that a supernova explosion represents a unique scenario
for further study of the PMNS matrix.
The corresponding  neutrinos, which carry  all flavors
were observed in only one occasion (SN1987A),
have an energy $E_\nu \lesssim 100$~MeV \cite{Str06}, and
are also studied through the interactions with
 carbon nuclei in the liquid scintillator.

Thus, the main interest in the neutrino/antineutrino-$^{12}$C
charge-exchange cross sections comes from the neutrino oscillations, and
 precise knowledge of the cross sections in the
neutrino energies  going from a  few  MeV´s up to a few GeV´s
is required.
Up to quite recently
the only available  experimental information on reactions was that for the
flux-averaged cross-sections:
 i) $^{12}$C$(\nu_e,e^-)^{12}$N
 in the DAR region: $E_{\nu_e} < 60$~MeV~\cite{Ath97,Aue01,Zei98},
 and ii) $^{12}$C$(\nu_\mu,\mu^-)^{12}$N in the  DIF region:
 $127$ MeV $\leq E_{\nu_\mu}\leq 300$ MeV~\cite{Ath97a,Aue02a,LSND}.
 In last few years, however, several
experimental programs at MiniBooNE~\cite{MiniBooNE Collaboration},
K2K~\cite{K2K Collaboration}, and  SciBooNE~\cite{SciBooNE Collaboration} yield
results on the ($\nu_\mu,^{12}$C)  cross section for
 $0.4$ GeV $\leq E_{\nu_\mu}\leq 1.7$ GeV.
 It is well known  that for $E_\nu$ larger than a few hundreds  MeV's, besides
 the quasi-elastic (QE) channel,
  many inelastic channels are open and pion production becomes important.
In fact, there have  been quite active experimental efforts to
investigate neutrino-induced coherent single-pion production
in the  $\Delta$-excitation region of $^{12}$C. Starting approximately at the  threshold
coming from the pion and charged lepton masses ($\mass_\pi$ and $\mass_\ell$), the $\pi+\ell$
 production cross section steadily increases with the neutrino energy becoming
  larger than the quasi-elastic one for $E_\nu\lsim 1.5$
  GeV~\cite{MiniBooNE Collaboration, K2K Collaboration,
SciBooNE Collaboration}.

From the  theoretical side there have been great efforts to
understand the  nuclear structure within the triad
$\{^{12}$B,$^{12}$C,$^{12}$N$\}$. In the seminal   work of
O'Connell,  Donelly,  and Walecka ~\cite{Con72} a unified
analysis of electromagnetic, and semileptonic weak interactions was presented. To
describe the nuclear dynamics they have used the particle-hole
Tamm-Dancof Approximation (TDA) within a very small single-particle space
\footnote{From now on a    single-particle (s.p.) space
that includes all orbitals  within  $N$ harmonic oscillator (HO)
shells will be  labeled as space $S_{N}$.}
($S_2$ $\equiv \{1s_{1/2}$, $1p_{3/2}$, $1p_{1/2}$, $1d_{5/2}$, $2s_{1/2}\}$)~\cite{Don70}.
To achieve  agreement with
experiments for the $\beta^\pm$-decays, and $\mu$-capture they were
forced to use an overall reduction factor $\xi^2$ of the order of
$4$ ($2$) for even (odd) parity states. They have also pointed out
that this factor would become totally unnecessary with  use of a
better nuclear model able to open the $1p_{3/2}$ shell.

Rather thorough comparisons of $2s1d$ and $2p1f$ shell-model predictions
with measured allowed $\beta$-decay rates have yielded a simple,
phenomenological effective axial coupling $g_{\sss
A}^{\sss}=1$ that should be used rather than the bare value
\cite{Bro85,Cas87,Ost92,Mar96}. This observation is the basis for
many nuclear model estimates of the  Gamow-Teller (GT) response that governs
allowed neutrino cross sections.
 In Ref.~\cite{Con72}
$g_{\sss A}^{\sss}=1.23$ was used based on a study of neutron
$\beta$-decay, and,  as the analyzed  processes were dominantly of the
axial-vector type, the use of $g_{\sss A}^{\sss}=1$ would  have
diminished the reduction factors $\xi^2$ in an appreciable way.

In the Random Phase Approximation (RPA),  besides the TDA forward-going
amplitudes,  the backward-going amplitudes are present as well. However,
these additional  RPA amplitudes did not help to open the
$1p_{3/2}$ shell in the continuum RPA (CRPA) calculations
of Kolbe, Langanke, and Krewald~\cite{Kol94}. Thus,  as in
the case of the TDA used  in Ref.~\cite{Con72},
to get agreement with data
for the ground state triplet $T=1$ ($\beta^\pm$-decays, $\mu$-capture,
and the exclusive $^{12}$C$(\nu_e,e^-)^{12}$N reaction)
their calculations were  rescaled  by a factor $\cong 4$.

The main aim of the CRPA is to  describe
appropriately not only the bound states but also  the virtual
(quasi-bound), resonant, and  continuum states,
which are treated as bound states in the RPA. However, this
superiority has not been evidenced so far in numerical
calculations.  For instance, in the case of $\mu$-capture rates
in $^{16}$N  the two methods agree with
each other quite well for the  $0^{-}$ and $1^{-}$  states,
while the RPA result is preferred
for the $2^{-}$  state ~\cite{Kol94a}.

To open the $1p_{3/2}$ shell one has to introduce pairing correlations.
This is done within the Shell Model (SM)~\cite{Hay00,Vol00,Suz06}, which
reproduces quite well both i) the
experimental flux-averaged exclusive,
and inclusive cross sections
for the $^{12} \rm C(\nu_e,e^-)^{12}\rm N$ DAR~\cite{Ath97,Aue01,Zei98}, and
$^{12}\rm C(\nu_\mu,\mu^-)^{12}\rm N$ DIF~\cite{Ath97a} reactions, and
ii) the $\mu^- + {^{12}\rm C}\rightarrow \nu_{\mu}+{^{12}\rm B}$
muon-capture modes~\cite{Mil72,Mea01,Sto02}.

The quasiparticle RPA (QRPA) also opens the $1p_{3/2}$ shell
by means of the pairing interaction. However,  it  fails as well in
 accounting for the exclusive processes
to the isospin triplet $T=1$ in $^{12}\rm C$, because
a new problem emerges, as  first
observed by Volpe \etal~\cite{Vol00}.   They noted that within the QRPA
the lowest state in ${^{12}\rm N}$ irremediably turned out not to be
the most collective one.  Later it was shown
\cite{Krm02,Krm05,Sam06} that: 1) the origin of this difficulty
arises from the degeneracy among  the four lowest proton-neutron
two-quasiparticle ($2qp$) states $\ket{1p_{1/2}1p_{3/2}}$,
$\ket{1p_{3/2}1p_{3/2}}$, $\ket{1p_{1/2}1p_{1/2}}$ and
$\ket{1p_{3/2}1p_{1/2}}$, which, in turn, comes from the fact that
for $N=Z=6$ the quasiparticle energies $E_{1p_{1/2}}$ and
$E_{1p_{3/2}}$
are very close to  each other, and 2) it is imperative to
use the projected QRPA (PQRPA) for a physically sound description
of  the weak processes among  the ground states of the triad
$\{{{^{12}\rm B},{^{12}\rm C},{^{12}\rm N}}\}$~\cite{Krm02,Krm05,Sam06};
see  Figs. 2 and 3 in Ref.~\cite{Krm05}.

In summary, neither the CRPA nor the QRPA are the appropriate
nuclear models to describe the
``fine structure" of exclusive  charge-exchange processes around $^{12}$C, and
they only can be used for global inclusive descriptions. Of course,
the same is valid for the relativistic RPA
(RQRPA) that  has recently been applied with success in
calculations of inclusive charged-current neutrino-nucleus reactions
in $^{12}$C, $^{16}$O, $^{56}$Fe, and  $^{208}$Pb~\cite{Paa07}, and
 total muon capture rates on a large set of nuclei
from $^{12}$C to $^{244}$Pu~\cite{Mar09}.
The continuum QRPA (CQRPA)
would have to be superior to the QRPA for the same reasons
that the CRPA  would have to be better than the RPA.
Nevertheless, neither this superiority  has been put
in evidence by  numerical calculations~\cite{Hag01,Rod08}.
Finally, it is clear
that the nuclear structure descriptions  inspired on the Relativistic
Fermi Gas Model (RFGM)~\cite{Smi72,Nie04,Val06}, which do not involve
multipole expansions,  should only  be used for inclusive
quantities.

When the effects due to resonant and continuum states are
considered, as it is done
within the CRPA and CQRPA,
the  spreading in strength
of the hole states in the inner shells should also be taken into account
 for the sake of consistency.
In fact, a single-particle state $j$ that is  deeply bound in the parent nucleus,  after
a weak interacting process  can become a highly excited hole-state
$j^{-1}$ in the continuum of the residual nucleus.
There it is suddenly
mixed with more complicated configurations
(2h1p, 3h2p, \dots excitations, collective states, and so on)
spreading its strength in a relatively wide energy interval~\cite{Ma85}\footnote{One should keep
in mind that the mean life of $^{12}$N and $^{12}$B are, respectively,
$11.0$ and $20.2$  ms, while  strong interaction times are
of the order of $10^{-21}$ s.}.
This happens, for instance, with the $1s_{1/2}$ orbital in $^{12}$C,
that is separated  from the $1p_{3/2}$ state by approximately $23$ MeV, which
is enough to break the $12$ particle system, where the energy of the last
excited state amounts to $ 11.5$ MeV in $^{12}$N, and $ 16.5$ MeV in $^{12}$B  channels.
Although  the detailed structure and fragmentation
of hole states are still not well known, the
exclusive knockout reactions provide a wealth of
information on the structure of single-nucleon states
of nuclei. Excitation energies  and widths of
proton-hole states were systematically measured with
quasifree (p, 2p) and $(e, e' $p) reactions, which
revealed the existence of inner orbital shells in
nuclei~\cite{Ja73,Fr84,Be85,Le94,Ya96,Ya01,Yo03,Ya04,Ko06}.

In the  TDA calculation of Ref.~\cite{Con72} the  $S_2$
space has been used, which  extends only
from $13.77$ MeV up to $30.05$ MeV,
embracing, respectively, $1, 2, 2, 1$, and $1$ negative parity
 states $J^\pi=0^-, 1^-, 2^-, 3^-$, and $4^-$,
and $1, 2, 2$, and $1$ positive parity
 states $J^\pi=0^+, 1^+, 2^+$, and $3^+$.
 With such  small configuration space,
the neutrino cross sections $\sigma_e(E_\nu)$, and  $\sigma_\mu(E_\nu)$  have been evaluated
up to a neutrino energy $E_\nu$
of $0.6$ GeV, and extrapolated up to $20$ GeV.
In recent years, however,  large configuration spaces
 have been used
in the evaluation of  QE cross sections for $E_\nu\sim 1$ GeV.
For instance,  Amaro \etal~\cite{Ama05}
have employed the single-particle SM (TDA without the residual interaction)
in a semirelativistic description of quasielastic neutrino
reactions $(\nu_\mu,\mu^-)$ on $^{12}$C going up to $E_\nu= 1.5$ GeV, and including
multipoles $J^\pi\leq 47^{\pm}$.
Good agreement with the RFGM was obtained for
several choices of kinematics of interest for the ongoing neutrino
oscillation experiments. Kolbe \etal~\cite{Kol03} have also achieved an excellent
agreement between the RFGM and the CRPA calculations of the total cross
 section and the angular distribution of the  outgoing electrons in $^{16}{\rm O}(\nu_e,e)X$
for $E_\nu\le  0.5$ GeV. They have considered  states  up to $J^\pi=9^{\pm}$ only,  and
didn't specify the configuration
space used. Moreover, Valverde \etal~\cite{Val06} have analyzed
the theoretical uncertainties of the RFGM developed in~\cite{Nie04}  for the $(\nu_e,e^-)$, and
 $(\nu_\mu,\mu^-)$ cross sections in  $^{12}$C, $^{16}$O,
and $^{40}$Ca for $E_\nu\le  0.5$ GeV.
The work of
 Kim \etal~\cite{Kim08} should also be mentioned
 where were studied the effects of strangeness
 on the $(\nu_\mu,\mu^-)$ and  $({\tilde\nu}_\mu,\mu^+)$ cross sections in  $^{12}$C
 for incident energies between  $0.5$ MeV and $1.0$ GeV,  within the
  framework of a relativistic single-particle model.
  Quite recently,
 Butkevich~\cite{But10} has also studied the
 scattering of muon neutrino on carbon targets for neutrino energies up to $2.8$ GeV
within  a relativistic shell-model approach without specifying
the model space.

For relatively large neutrino energies ($E_{\nu_e}\gsim \mass_\pi$, and
$E_{\nu_\mu}\gsim \mass_\pi+\mass_\mu$)
to the above-mentioned
 QE cross sections  should be added the pion production cross section,  as done, for instance,
 in Refs.~\cite{Mart09,Lei09}.
One should also note that
$\sigma_e(E_{\nu_e})$ and $\sigma_\mu(E_{\nu_\mu})$ coincide with each other asymptotically.
This is clearly put in evidence in  the Extreme Relativistic
Limit (ERL) where $|\pb_\ell|/E_\ell\go 1$, and the neutrino-nucleus cross sections
depend on $\mass_\ell$ only trough the threshold energy, as can be seen
from the Appendix~\ref{C} of the present work. The  Figure 4 from Ref.~\cite{Kur90} is
also illustrative in this respect.

 Therefore, we  focus our attention only
 on the quasi-elastic cross section $\sigma_e(E_{\nu_e})$ since,
 at muon-neutrino energies involved in the MiniBooNE experiment~\cite{MiniBooNE}, it
 is equal to $\sigma_\mu(E_{\nu_\mu})$   for all practical purposes.

One of main objectives in the present study is to analyze the effect of the size of
the configuration space
up to neutrino energies of several hundred MeV.
As in several previous works
\cite{Con72,Mar96,Kol94,Kol94a,Hay00,Vol00,Suz06,Krm02,Krm05,Sam06,Paa07,Mar09,
Ama05,Kol03,Val06,Kim08,Kur90} this will be done in first order perturbation theory.
The consequences  of the particle-particle force in the S = 1, T = 0 channel,
within the PQRPA will also  be examined.  The importance of this piece of the residual
interaction was discovered more than 20 years ago by  Vogel and Zirnbauer~\cite{Vog86}
and Cha~\cite{Cha87}, and since then
the QRPA  became  the most frequently used nuclear structure
method for evaluating double  $\beta$-decay rates.

A few words will be devoted as well as to the non-relativistic
formalisms for neutrino-nucleus scattering.
The most popular  one was developed by  the Walecka group~\cite{Con72,Don79,Hax79,Wal95},
where the
nuclear transition matrix elements are classified as Coulomb,
longitudinal, transverse electric,
and transverse magnetic multipole moments. We feel that these
denominations might be convenient
when  discussing simultaneously  charge-conserving, and
charge-exchange processes, but seems  unnatural when
one considers  only the last ones.
 As a matter of fact, this terminology is not often used in  nuclear
$\beta$-decay,  $\mu$-capture, and charge-exchange reactions where  one only
speaks   of vector and axial matrix elements with different
degrees of   forbiddenness:
allowed (GT and Fermi), first forbidden, second forbidden, \etc., types~\cite{Beh82,Krm80}.
There are exceptions, of course, as
for instance, is the recent work of Marketin \etal~\cite{Mar09} on muon capture,
where  the Walecka's classification  was used.

The formalism
worked out by  Kuramoto~\cite{Kur90}  is also frequently used for the evaluation
  of neutrino-nucleus cross-sections. It is   simpler than that of Walecka,
  but it does not contain   relativistic matrix elements, nor  is applicable
for  muon capture rates.

More recently,
we have introduced another formalism~\cite{Krm02,Krm05,Sam06}.
  Besides of being   almost as simple as that of Kuramoto, it
retains  relativistic terms and can be used for $\mu$-capture.
This formalism is briefly sketched here, including
the consequences of the violation of the Conserved Vector Current
(CVC) by the Coulomb field.
It is furthermore simplified by  classifying the nuclear matrix
elements in natural and
unnatural parities.
We also show how within the present formalism both the sum rule approach,
 and the formula for ERL  look like.

In Section \ref{Sec2} we briefly describe  the formalism used to evaluate different  weak interacting
processes. Some details are delegated to the Appendices: \ref{A}) Contributions of
natural and unnatural parity
states to the transition rates, \ref{B}) Sum rule approach for the inclusive
neutrino-nucleus cross section,
\ref{C}) Formula for the inclusive neutrino-nucleus cross section at the extreme relativistic limit, and
\ref{D}) Formula for the muon capture rate.
  In Section \ref{Sec3}  we present, and discuss the numerical results.
  Comparisons with experimental data, as well as with
previous theoretical studies, are  done   whenever possible.
Here we firstly sketch the two theoretical frameworks, namely
the PQRPA and RQRPA,  used in the
numerical calculations. In
subsections \ref{Sec3A}, and \ref{Sec3B}  we present the results
for the exclusive and inclusive processes, respectively.
Finally, in Section \ref{Sec4} we give a brief summary, and final conclusions.

\section{Formalism for the Weak Interacting Processes}
\label{Sec2}
The weak Hamiltonian is expressed in the form~\cite{Don79,Wal95,Bli66}
\br
H_{{\sss {W}}}(\rb)&=&\frac{G}{\sq}J_\alpha l_\alpha e^{-i\rb\cdot\kb},
\label{2.1}\er
where $G=(3.04545\pm 0.00006){\times} 10^{-12}$ is the Fermi coupling
constant (in natural units), the leptonic current $l_\alpha\equiv
\{ \lb,il_\emptyset\}$ is given by the Eq. (2.3) in Ref. \cite{Krm05} and
 the hadronic current operator $J_\a\equiv
\{ \Jb,i J_\emptyset \}$ in its nonrelativistic form reads
\footnote{ As in Ref. \cite{Krm05} we use the Walecka's notation
\cite{Wal95} with the Euclidean metric for the quadrivectors,
and $\alpha=1,2,3,4$. The only difference is that we substitute
his indices $(0,3)$ by our indices $(\emptyset,0)$, where we use
the index $\emptyset$ for the temporal component and the
index 0 for the third spherical component.}
\br
 J_\emptyset&=&\gV + (\ga+\gpa) {\mbs}\cdot\hat{\kb} +\gas\frac{i \mbs \cdot\mbn}{\rm M},
\label{2.2}\\
{ {\bf J}}&=& -\gA {\mbs} -i \gw {\mbs}\x\hat{\kb}-\gv \hat{\kb}
+\gpb({\mbs} \cdot\hat{\kb})\hat{\kb}-\gvs \frac{i \mbn}{\rm M},
\nonumber\er
where $\hat{\kb}\equiv{\kb}/{|\kb|}$.
The quantity \br k = P_i-P_f\equiv \{\kb,ik_\emptyset
\} \label{2.3}\er is the momentum transfer, ${\rm M}$ is the
nucleon mass, and $P_i$ and $P_f$ are momenta of the initial and
final nucleon (nucleus). The effective  vector, axial-vector,
weak-magnetism and pseudoscalar dimensionless coupling constants
are, respectively~ \br g_{\sss V}&=&1 , ~g_{\sss A}=1 , ~g_{\sss
M}=\kappa_p-\kappa_n=3.70,
\nn\\
~g_{\sss P}&=&g_{\sss A}\frac{2\Mass \mass_\ell }{k^{2}+\mass_\pi^2},
\label{2.4}\er
where
the following short notation has been introduced:
\br
\gv&=&\gV\frac{\k}{2\Mass};~
\ga=\gA\frac{\k}{2\Mass};~
\gw=(\gV+\gM)\frac{\k}{2\Mass},
\nn\\
\gpa&=&\gP\frac{\k}{2\Mass}\frac{k_\emptyset}{\mass_\ell};~
\gpb=\gP\frac{\k}{2\Mass}\frac{\k}{\mass_\ell}, \label{2.5}\er
 where
$\k\equiv |\kb|$. The above  estimates for $g_{\sss M}$ and
$g_{\sss P}$ come from the conserved vector current (CVC)
hypothesis, and from the partially conserved axial vector current
(PCAC) hypothesis,  respectively.  The finite nuclear size (FNS) effect is
incorporated via the dipole form factor with a cutoff
$\Lambda=850$ MeV, i.e., $ g\go g\left[
{\Lambda^{2}}/(\Lambda^{2}+k^{2})\right]^{2}$.

In performing the multipole expansion of the nuclear operators
\be
O_\alpha\equiv (\Ob,iO_\emptyset)=J_\alpha e^{-i\kr},
\label{2.6}\ee
it is convenient 1) to take the momentum $\kb$  along  the
$z$ axis, i.e.,
 \br
 e^{-i\kr}&=&\sum_{\sf L}i^{-\sf L}\sqrt{4\pi(2{\sf L}+1)}
 j_{\sf L}(\rho) Y_{{\sf L}0}(\hat{\rb}),
\nn\\
 &=&\sum_{ \sf J}i^{-\sf J}\sqrt{4\pi(2{\sf J}+1)}
 j_{ \sf J}(\rho) Y_{{\sf J}0}(\hat{\rb}),
\label{2.7}\er
where $\rho=\k r$, and 2) to define  the operators
${\sf O}_{\a}$ as
 \be
  O_\alpha=\sqrt{4\pi}\sum_{ \sf J}i^{-{\sf J}}\sqrt{2{\sf J}+1}{\sf O}_{\a{\sf J}}.
\label{2.8}\ee
In this way we avoid the troublesome
factor $i^{-{\sf J}}$. In spherical coordinates (${m}=-1,0,+1)$ we
have
 \br
 J_\emptyset&=&\gV + (\ga+\gpa) \sigma_0  +i\gas{\rm M}^{-1}\mbs \cdot\mbn
\nn\\
J_{m}&=& -\gA \sigma_{m}+{m}\gw\sigma_{m}+\delta_{{m}0}[-\gv+\gpb\sigma_{0}]
\nn\\
&-&i\gvs{\rm M}^{-1}\nabla_m ,
\label{2.9}\er
and
 \br
 {\sf O}_{\emptyset{\sf J}}&=&
j_{\sf J}(\rho)Y_{{\sf J}0}(\hat{\rb})  J_\emptyset,
\nn\\
{\sf O}_{{m}{\sf J}}&=&\sum_{{\sf L}}i^{ \sf J-L}F_{{\sf
LJ}m}j_{\sf L}(\rho) \left[Y_{{\sf L}}(\hat{\rb})\otimes{\bf
J}\right]_{\sf J}, \label{2.10}\er where \br F_{{\sf
LJ}m}&\equiv&(-) ^{\sf J+ m}\sqrt{2{\sf L}+1} \threej{{\sf
L}}{1}{{\sf J}}{0}{-m}{m}
\nn\\
&=&(-) ^{1+ m}(1,-m{\sf J}m|{\sf L}0),
\label{2.11}\er
is a Clebsch-Gordan coefficient.
\footnote{Their values are:
\brn
F_{{\sf J}+1,{\sf J},0}&=&-\sqrt{\frac{{\sf J}+1}{2{\sf J}+1}},~~~~
F_{{\sf J}-1,{\sf J},0}=\sqrt{\frac{{\sf J}}{2{\sf J}+1}},
\nn\\
F_{{\sf J}+1,{\sf J},\pm 1}&=&\sqrt{\frac{{\sf J}}{2(2{\sf J}+1)}},~~~~F_{{\sf J},{\sf J}-1,\pm 1}=\sqrt{\frac{{\sf J}+1}{2(2{\sf J}+1)}},
\nn\\
F_{{\sf J},{\sf J},0}&=&0,~~~~F_{{\sf J},{\sf J},\pm 1}=\mp{1\over \sqrt{2}}.
\ern}

Explicitly, from \rf{2.9} 
\br {\sf O}_{\emptyset{\sf
J}}&=&g_{\sss{V}}\M_{\sf J}^{\sss V} +i\gas\M^{\sss A}_{\sf
J}+i(\ga+\gpa)\M^{\sss A}_{0{\sf J}},
\label{2.12}\\
{\sf O}_{{m}{\sf J}} &=&i(\delta_{{m}0}\gpb-\gA +m \gw)\M^{\sss
A}_{{m}{\sf J}}
\nn\\
&+&\gvs\M^{\sss V}_{{m}{\sf J}}-\delta_{{m} 0}\gv\M_{\sf J}^{\sss V}.
\label{2.13}\er
The elementary operators are given by
\br
\M^{\sss V}_{\sf J}&=&j_{\sf J}(\rho) Y_{{\sf J}}(\hat{\rb}),
\nn\\
\M^{\sss A}_{\sf J}&=&
{\rm M}^{-1}j_{\sf J}(\rho)Y_{\sf J}(\hat{\rb})(\mbs\cdot\mbn),
\label{2.14}\er
and
\br
\M^{\sss A}_{{m\sf J}}&=&\sum_{{\sf L}\ge 0}i^{ {\sf J-L}-1}
F_{{{\sf LJ}m}}j_{\sf L}(\rho)
\left[Y_{{\sf L}}(\hat{\rb})\otimes{\mbs}\right]_{{\sf J}},
\nn\\
\M^{\sss V}_{{m\sf J}}&=&{\rm M}^{-1}\sum_{{\sf L}\ge 0}i^{ {\sf J-L}-1}
F_{{{\sf LJ}m}}j_{\sf L}(\rho)
[ Y_{\sf L}(\hat{\rb})\otimes\mbn]_{{\sf J}}.
\nn\\\label{2.15}\er

The CVC  relates the vector-current pieces of the operator \rf{2.6}
as (see Eqs. (10.45) and (9.7) from Ref.~\cite{Beh82})
\br
\kb\cdot\Ob^{\sss V}&\equiv&\k O_0^{\sss V}=
\tilde{k}_{\emptyset}O_\emptyset^{\sss V},
 \label{2.16}\er
 with
\br
\tilde{k}_{\emptyset}&\equiv& k_{\emptyset}
-S(\Delta E_{\rm Coul}-\Delta\Mass),
 \label{2.17}\er
where
\br
\Delta E_{\rm Coul}\cong \frac{6e^2Z}{5R} \cong 1.45 ZA^{-1/3}~~\mbox{MeV},
\label{2.18}\er
is the Coulomb energy difference between the initial and final nuclei,
$\Delta\Mass=\Mass_n-\Mass_p=1.29$ MeV is the neutron-proton
mass difference, and $S=\pm 1$ for neutrino and antineutrino scattering, respectively.

The consequence of the CVC relation \rf{2.16} is  the substitution
\br \gvs\M^{\sss V}_{{\sf 0}{\sf J}}-\gv\M_{\sf J}^{\sss V}\go
\frac{\tilde{k}_{\emptyset}}{\k}g_{\sss{V}} \M^{\sss V}_{\sf J},
\label{2.19}\er
 in  \rf{2.13}, and ${\sf O}_{{m}{\sf J}}$ now reads
\br {\sf O}_{{m}{\sf J}} &=&i(\delta_{{m}0}\gpb-\gA +m \gw)
\M^{\sss A}_{{m}{\sf J}}
\nn\\
&+&|m|\gvs\M^{\sss V}_{{m}{\sf J}}+\delta_{{m} 0}
\frac{\tilde{k}_{\emptyset}}{\k}\gV\M_{\sf J}^{\sss V}.
\label{2.20}\er

The second term in \rf{2.17}
comes from the  violation of the CVC  by the electromagnetic interaction.
Although it is frequently employed   in the nuclear $\b$-decay,
as far as we know, it has never been considered  before in the
neutrino-nucleus scattering. $\Delta E_{\rm Coul}$ is equal
to $3.8, 9.8$, and $20.0$ MeV for $^{12}$C, $^{56}$Fe, and $^{208}$Pb,
respectively, and therefore the just mentioned term  could be quite significant,
specially for  heavy nuclei.

The  transition amplitude for the neutrino-nucleus reaction at
a fixed value of $\k$,
from the initial state  $\ket{0^+}$ in the $(Z,N)$ nucleus to
the n-th final state $\ket{{\sf J}^\pi_n}$ in the
nucleus $(Z\pm 1,N\mp 1)$,  reads
\br
\T_{{\sf J}^\pi_n}(\k)\equiv
\sum_{ s_\ell,s_\nu }
\left|\bra{{\sf J}_n^\pi }H_{{\sss {W}}}(\k)\ket{0^+}\right|^{2}.
\label{2.21}\er
The momentum transfer here is
$k=p_\ell-q_\nu$, with
 $p_\ell\equiv\{\pb_\ell,iE_\ell\}$
and $q_\nu\equiv\{\qb_\nu,iE_{\nu}\}$, and after some
algebra~\cite{Krm05} one gets
\br \T_{{\sf J}^\pi_n}(\k)&=&{4\pi
G^2} [\sum_{\a=\emptyset,0,\pm 1} |\Bra{{\sf J}^\pi_n}{\sf O}_{\a{
{\sf J}}}(\k)\Ket{0^+}|^2 \L_{\a}
\label{2.22}\\
&-&2\Re\left(\Bra{{\sf J}^\pi_n}{\sf O}_{\emptyset{ {\sf J}}}(\k)
\Ket{0^+} \Bra{{\sf J}^\pi_n}{\sf O}_{0{ {\sf
J}}}(\k)\Ket{0^+}^*\right) \L_{\emptyset 0}], \nn\er
 where \br
\L_{\emptyset}&=&1+\frac{|\pb_\ell|\cos\theta}{E_\ell},
\nn\\
\L_{0}&=&1+\frac{2q_0p_0}{E_\ell
E_\nu}-\frac{|\pb_\ell|\cos\theta}{E_\ell},
\nn\\
\L_{\pm1}&=&1-\frac{q_0p_0}{E_\ell E_\nu}\pm
\left(\frac{q_0}{E_\nu}-\frac{p_0}{E_\ell}\right)S ,
\nn\\
\L_{\emptyset 0}&=&\frac{q_0}{E_\nu}+\frac{p_0}{E_\ell},
\label{2.23}\er
are the lepton traces, with $\theta\equiv \hat{\qb}_\nu\cdot\hat{\pb}_\ell$
being the angle between
the incident neutrino and ejected lepton momenta, and
\br
q_0&=&{\hat k}\cdot\qb_\nu=\frac{E_\nu(|\pb_\ell|\cos\theta-E_\nu)}{\k},
\nn\\
p_0&=&{\hat k}\cdot \pb_\ell=
\frac{|\pb_\ell|(|\pb_\ell|-E_\nu\cos\theta)}{\k},
\label{2.24} \er
 the $z$-components of the neutrino and lepton momenta.

The exclusive  cross-section (ECS) for the state $\ket{{\sf J}^{\pi}_n}$, as a function of the
incident neutrino energy $E_{\nu}$, is
\br
\s_\ell({\sf J}^{\pi}_n,E_{\nu})& = &\frac{|\pb_\ell|
E_\ell}{2\pi} F(Z+S,E_\ell)
\int_{-1}^1
d(\cos\theta)\T_{{\sf J}^\pi_n}(\kappa),
\nn\\
\label{2.25}\er
where
\br
E_\ell&=&E_\nu-\w_{{\sf J}^\pi_n},~
|\pb_\ell|=\sqrt{(E_{\nu}-\w_{{\sf J}^\pi_n})^2-m_\ell^2},
\nn\\
\kappa&=&|\pb_\ell-\qb_{\nu}|\nn\\
&=&\sqrt{2E_{\nu}(E_\ell-|\pb_\ell|\cos\theta)-m_\ell^2
+\w_{{\sf J}^\pi_n}^2},
\label{2.26}\er
and $\w_{{\sf J}^\pi_n}=-k_{\emptyset}=E_\nu-E_\ell$
 is the excitation
energy of the  state $\ket{{\sf J}^\pi_n }$ relative to
the  state $\ket{0^+}$.
Moreover,  $F(Z+S,E_\ell)$ is  the Fermi function for neutrino $(S=1)$,
and antineutrino $(S=-1)$, respectively.

As it well known the charged-current cross sections must
be corrected for the distortion of the outgoing lepton wave
function by the Coulomb field of the daughter nucleus.
For relatively low neutrino energies ($\sim 40-50$ MeV)
this correction was implemented by numerical
solution of the Dirac equation for an extended nuclear charge \cite{Beh82}.
At higher energies, the effect of the Coulomb field was described by
the effective momentum approximation (EMA) \cite{Eng98},
in which the lepton momentum $p_\ell$ and energy $E_\ell$ are modified
by the corresponding effective quantities  (see \cite[Eq. (34) and (45)]{Paa07}).

Here, we will also deal with inclusive cross-section (ICS),
\br
\s_\ell(E_{\nu})=\sum_{{\sf J}^{\pi}_n}\s_\ell({\sf J}^{\pi}_n,E_{\nu}),
\label{2.27}\er
as well as with folded  cross-sections, both exclusive,
\br
\overline{\s}_\ell({{\sf J}^{\pi}_n})=\int dE_{\nu}\s_\ell({\sf J}^{\pi}_n,E_{\nu})
 n_\ell(E_{\nu}),
\label{2.28}\er
and inclusive
\be
\overline{\s}_\ell= \int dE_{\nu}
\s_\ell(E_\nu) n_\ell(E_{\nu}),
\label{2.29}\ee
where $n_\ell(E_{\nu})$ is the neutrino~(antineutrino) normalized flux.
In  the evaluation of both neutrino, and antineutrino ICS
the summation in \rf{2.27} goes
  over all $n$ states with spin and parity
  ${\sf J}^{\pi}\le 7^{\pm}$  in the PQRPA, and
  over ${\sf J}^{\pi}\le 14^{\pm}$ in the RQRPA.

In the Appendix \ref{A}  we show that the  real and imaginary
parts of the operators ${\sf O}_{{\a}{\sf J}}$, given by \rf{2.12}
and \rf{2.20}, contribute to  natural and unnatural parity states,
respectively. This greatly simplifies the numerical calculations,
because one always  deals with real operators only.
In Appendix \ref{D} are also shown the formula
for the muon capture process within the present formalism.

\begin{figure}[t]
{
\includegraphics[width=8.6cm,height=10.cm]{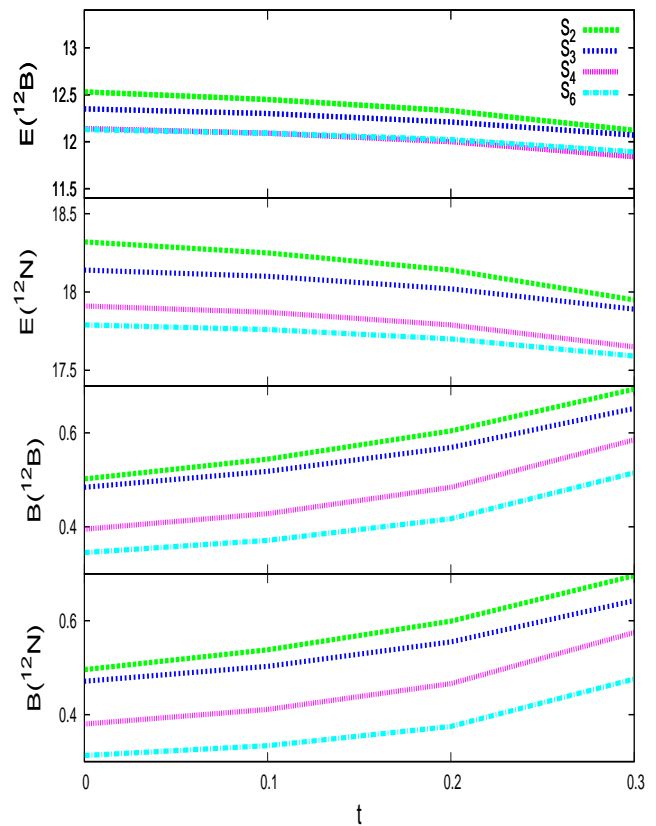}
\vspace{0.5cm}
\caption{\label{F1}  (Color online) $^{12}$B and $^{12}$N ground
state energies (in units of MeV) and GT $B$-values within the
PQRPA for different s.p. spaces, as function of the $pp$-coupling
$t$. The experimental values are: $E(^{12}$B$)= 13.37$ MeV, and $
E(^{12}$N$)=17.33$ MeV  ~\cite{Ajz85}, and $B(^{12}$B$)=0.466$,
and $B(^{12}$N$)=0.526$~\cite{Al78}. }
}
\end{figure}

\label{Sec3A}
\begin{figure}[h]
\begin{center}
{\includegraphics[width=8.6cm,height=12.cm]{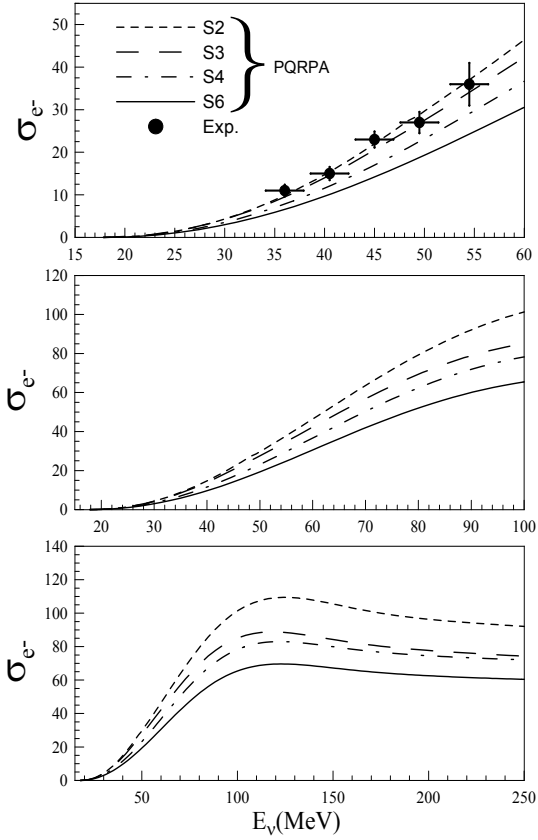}}
\end{center}
\vspace{-1cm}
\caption{\label{F2} Exclusive $^{12}$C($\nu,e^-)^{12}$N cross-section
$\sigma_e(E_\nu,1^+_{1})$ (in units of $10^{-42}$ cm$^2$),
plotted as a function of the incident neutrino
energy $E_\nu$. Results for several single-particle  spaces $S_N$, and $t=0$, within three  different
 energy intervals, are shown.
The experimental data in  the DAR region are from Ref.~\cite{Ath97}.}%
\end{figure}
\begin{figure}[h]
\begin{center}
{\includegraphics[width=8.6cm,height=12.cm]{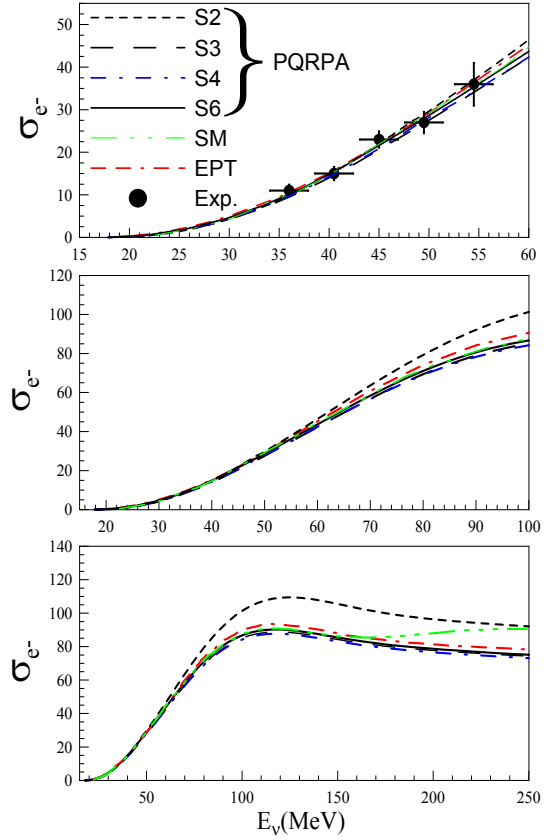}}
\end{center}
\vspace{-1cm}
\caption{\label{F3}(Color online)
Same as in Figure~\ref{F2}, but now $t=0$ for $S_2$, and  $S_3$, $t=0.2$ for $S_4$, and $t=0.3$ for $S_6$.
The SM, and EPT calculations are, respectively, from Refs.~\cite{Eng96}, and ~\cite{Fuk88}.
The experimental data in  the DAR region are from Ref.~\cite{Ath97}.}%
\end{figure}

\section{Numerical results and discussion}
\label{Sec3}
The major part of the numerical calculations have been done within the PQRPA
by  employing  the $\delta$-interaction (in MeV fm$^3$)
\[
V=-4 \pi \left(v_sP_s+v_tP_t\right) \delta(r),
\]
with singlet ($v_s$), and triplet ($v_t$) coupling constants
  different for the particle-hole ($ph$), particle-particle ($pp$), and pairing ($pair$)
channels~\cite{Sam08}.
This interaction leads to a good description of single and
double $\beta$-decays and it has been used extensively
in the literature~\cite{Hir90a,Krm92,Krm93,Krm94}.
The single-particle wave
functions were approximated with those of the HO with the
length parameter $b=1.67$ fm, which corresponds
to the oscillator energy $ \hbar \omega=45A^{-1/3}-25A^{-2/3}$~ MeV.
The  s.p. spaces $S_{2},S_{3},S_{4}$, and $S_{6}$
  will be explored.

In Refs.~\cite{Krm02,Krm05}, where  the $S_{3}$ space was used,  we have pointed out
that the values of the coupling strengths $v^{pp}_s$, $v^{pair}_s$, and $v^{p}_t$  used
in $N > Z$ nuclei ($v^{pp}_s=v^{pair}_s$,
$v^{pp}_t\gtrsim v^{pp}_s$),  might not be suitable for
 $N=Z$ nuclei.
 In fact, the best agreement with data in $^{12}$C  is obtained for: i) the
energy of the ground state in $^{12}$N, $E(^{12}$N), ii) the GT $B$-values
in  $^{12}$C, $ B(^{12}$N) and $B(^{12}$B),
and iii) the exclusive muon capture in
$^{12}$B, $\Lambda^{\rm exc}\equiv \Lambda^(1^+_1)$,  is obtained when the $pp$
channel is totally switched  off,  i.e., $v^{pp}_s\equiv v^{pp}_t=0$.
The adopted $ph$ coupling strengths are
$v^{ph}_s=27$  MeV~ fm$^3$ and  $v^{ph}_t=64$ MeV~ fm$^3$~\cite{Krm02}.
For the $pp$ channel it is convenient to define the parameters
\[
s=\frac{v_s^{pp}}{v_s^{pair}},\hspace{0.6cm}t=\frac{v_t^{pp}}{v_s^{pair}},
\]
where $v_s^{pair}= (v_s^{pair}(p)+v_s^{pair}(n))/2$~\cite{Krm94}.
As in our previous work on $^{12}$C,  we will use here the same singlet and
triplet $pp$ couplings,  i.e., $s\equiv t$~\cite{Krm02,Krm05}.
The states with ${\sf J}^\pi=0^+$, and ${\sf J}^\pi=1^+$ only depend on
$s$, and $t$, respectively, while all remaining  depend on both coupling strengths.

 The  s.p. energies and pairing
strengths for $S_2$, $S_3$, and $S_4$ spaces, were varied in a $\chi^2$ search to account
for the experimental spectra of odd-mass nuclei $^{11}$C, $^{11}$B, $^{13}$C, and $^{13}$N,
as explained in Ref.~\cite{Krm05}. This method, however, is not
practical for the space
 $S_6$ which comprises $21$ s.p. levels. Therefore in this case the energies were derived
 in the way done in Ref.~\cite{Paa07}, while the pairing strengths
were adjusted to reproduce the experimental
gaps in $^{12}$C~\cite{Sam09}, considering all the  quasiparticle energies up to
 $100$ MeV.

For the purpose of the present study, we also employ the RQRPA theoretical framework
~\cite{PNVR.04}. In this case the ground state is calculated in
the Relativistic Hartree-Bogoliubov model (RHB) using effective
Lagrangians with density dependent meson-nucleon couplings and DD-ME2
parameterization~\cite{LNVR.05}, and pairing correlations are described by
the finite range Gogny force~\cite{BGG.91}. Details of the formalism can
be found in Refs.~\cite{Paa.03,PVKC.07}.  The RHB equations, and
respective equations for mesons are usually solved by expanding the Dirac
spinors and the meson fields in a spherical harmonic oscillator basis
with $S_{20}$ s.p. space.
 In the present study of neutrino-nucleus cross sections,
with energies of incoming neutrinos up to $600$ MeV, we extend the number of
oscillator shells up to  $N=30$ in order to accommodate s.p.
states at higher energies necessary for description of
cross sections involving higher energies of incoming (anti)neutrinos.
The number of $2qp$ configurations in the RQRPA is constrained by the
maximal excitation energy $E_{2qp}$.
 Within the RHB+RQRPA framework  the oscillator basis is
used only in RHB to determine the ground state and single-particle spectra.
The resulting wave functions are converted to
coordinate space for evaluation of the RQRPA matrix elements.
However, it is the original HO basis employed in RHB that
determines the maximal $E_{2qp}$ and  the size of
RQRPA configuration space.

\subsection{Weak interaction properties of $^{12}$N and  $^{12}$B ground states}

Let us first compare the QRPA and  PQRPA within the smallest
 configuration space  $S_2$, which contains  $16$ ${\sf J}^\pi=1^+$ states,
and with null $pp$ coupling: $t=0.$
The PQRPA ground state energies in $^{12}$N, and  $^{12}$B,
 are, respectively: $\omega_{+1}(1^+)=18.319$ MeV, and $\omega_{-1}(1^+)=12.528$ MeV,
while the corresponding wave functions read
\br
\ket{^{12}{\rm
N}}&=&0.963\ket{1p^\pi_{3/2}1p^\nu_{1/2}}
+0.232\ket{1p^\pi_{3/2}1p^\nu_{3/2}}
\nn\\
&+&0.122\ket{1p^\pi_{1/2}1p^\nu_{3/2}}
+0.105\ket{1p^\pi_{1/2}1p^\nu_{1/2}}
\nn\\
&+&\cdots,\label{3.1}\er
and
 \br
  \ket{^{12}{\rm B}}&=&-0.971\ket{1p^\pi_{1/2}1p^\nu_{3/2}}
+0.204\ket{1p^\pi_{3/2}1p^\nu_{3/2}}
\nn\\
&-&0.125\ket{1p^\pi_{3/2}1p^\nu_{1/2}}
+0.090\ket{1p^\pi_{1/2}1p^\nu_{1/2}}
\nn\\
&+&\cdots\label{3.2}\er

The analogous QRPA   energies are quite similar: $\omega_{+1}(1^+)=17.992$ MeV,
 $\omega_{-1}(1^+)=12.437$ MeV. However, the wave functions are quite different.
The main difference is in the fact that QRPA  furnishes  the same wave functions
 for all four nuclei  $^{12}$N, $^{10}$B, $^{14}$N, and  $^{12}$B, being that of the
 ground state:
\br
\ket{1^+_{GS}}&=&-0.272\ket{1p^\pi_{3/2}1p^\nu_{1/2}}
-0.759\ket{1p^\pi_{3/2}1p^\nu_{3/2}}
\nn\\
&+&0.356\ket{1p^\pi_{1/2}1p^\nu_{3/2}}
-0.472\ket{1p^\pi_{1/2}1p^\nu_{1/2}}
\nn\\
&+&\cdots.\label{3.3}\er

The difference in the wave functions is an important issue  that
clearly signalizes towards the need for  the number
projection. In fact, the
PQRPA yields the correct limits ($1p^\pi_{3/2}\go 1p^\nu_{1/2}$ and
$1p^\nu_{3/2}\go 1p^\pi_{1/2}$) for   one-particle-one-hole (1p1h) excitations
on the $^{12}$C ground state to reach the $^{12}$N, and $^{12}$B nuclei.
 All
remaining configurations  in \rf{3.1}, and \rf{3.2} come from the higher order 2p2h, and
3p3h excitations. Contrary, the QRPA  state \rf{3.3} is
dominantly the two-hole excitation
$[(1p^\pi_{3/2})^{-1},(1p^\nu_{3/2})^{-1}]$, which corresponds to
the ground state of $^{10}$B.  More details
on this question can be found in Figure 3 of Ref.~\cite{Krm05}. The 1p1h
amplitudes$[(1p^\pi_{3/2})^{-1},1p^\nu_{1/2}]$, and
$[(1p^\nu_{3/2})^{-1},(1p^\pi_{1/2})]$  are dominantly present in
the following  QRPA states
 \br
\ket{1^+_{2}}&=&0.708\ket{1p^\pi_{1/2}1p^\nu_{3/2}}
+0.703\ket{1p^\pi_{3/2}1p^\nu_{1/2}}
\nn\\
&+&\cdots\
\nn\\
\ket{1^+_{4}}&=&-0.476\ket{1p^\pi_{3/2}1p^\nu_{1/2}}
+0.437\ket{1p^\pi_{3/2}1p^\nu_{3/2}}
\nn\\
&+&0.441\ket{1p^\pi_{1/2}1p^\nu_{3/2}}
-0.096\ket{1p^\pi_{1/2}1p^\nu_{1/2}}
\nn\\
&+&\cdots\label{4.12}\er
\begin{figure}[th]
\begin{center}
\hspace{-1.cm}
{\includegraphics[width=8.6cm,height=12.cm]{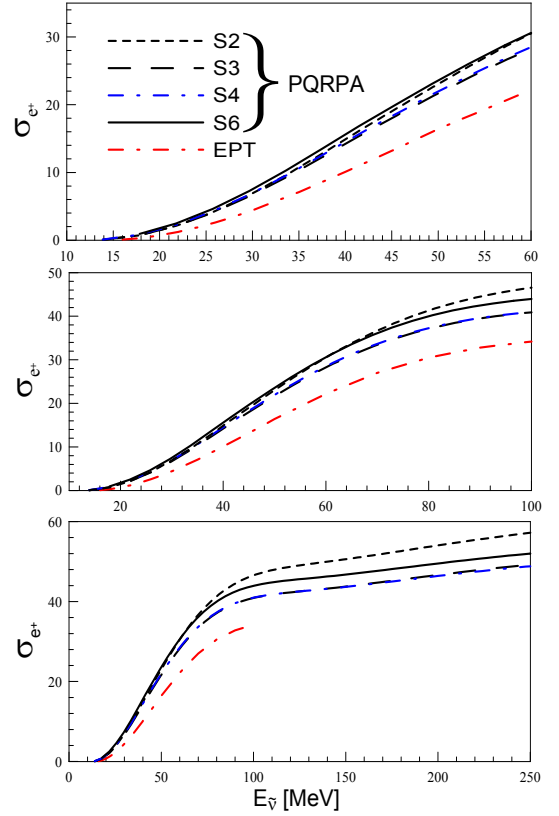}}
\end{center}
\vspace{-1cm}
\caption{\label{F4}(Color online) The calculated $^{12}$C(${\tilde{\nu}},e^+)^{12}$B cross-section
 $\sigma_{e^+}(E_{\tilde{\nu}},1^+_{1})$ (in units of $10^{-42}$ cm$^2$),
plotted as a function of the incident antineutrino
energy $E_{\tilde{\nu}}$. Same as in Figure ~\ref{F3}, the value of $t$ is $0$
for the s.p. spaces $S_2$, and  $S_3$, $0.2$ for  $S_4$ and $0.3$ for $S_6$.
The  EPT calculation from Ref.~\cite{Fuk88} is also shown.}%
\end{figure}

The  wave functions displayed above clearly evidence the
superiority of the PQRPA on the QRPA. Therefore from now on only the
PQRPA results will be discussed for the exclusive observables.

In Figure ~\ref{F1} we show  the $^{12}$B and $^{12}$N ground state
energies, and the corresponding GT $B$-values within the PQRPA for
different s.p. spaces, as function of the $pp$-coupling  $t$.
One sees that the
energies depend rather weakly on both, and  agree fairly well with the measured energies: $E(^{12}$B$)= 13.37$ MeV, and $
E(^{12}$N$)=17.33$ MeV  ~\cite{Ajz85}, although
the first one is somewhat underestimated, while  the second one is somewhat overestimated.
Both GT $B$-values  significantly increase with $t$ and diminish when size of
the s.p. space is increased.
For spaces $S_2$ and $S_3$ the best overall agreement with data  ($B(^{12}$B$)=0.466$,
and $ B(^{12}$N$)=0.526$~\cite{Al78}) is achieved
with $t=0$, while for spaces $S_4$ and $S_6$ this happens when the couplings are, respectively,
$t=0.2$, and $t=0.3$.
\begin{figure}[h]
\includegraphics[width=8.6cm,height=10.cm]{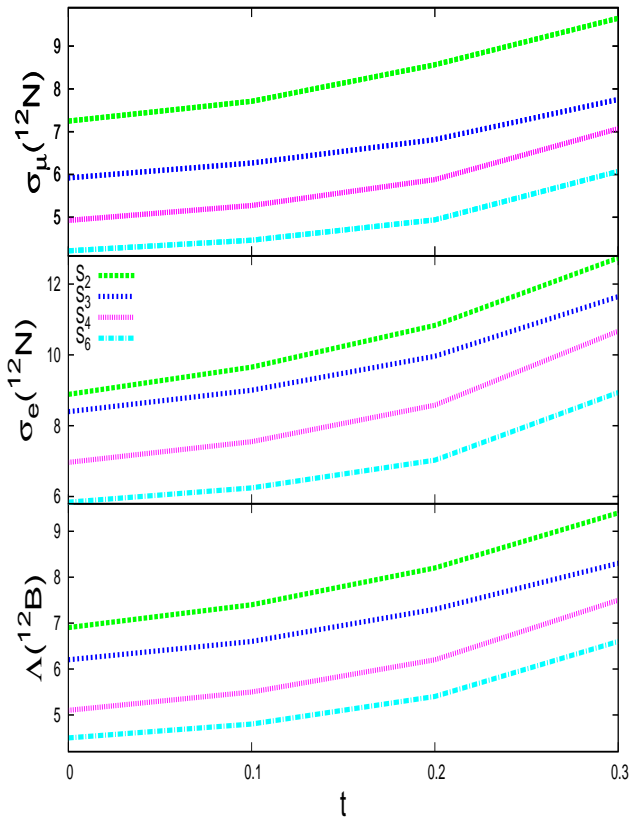}
\vspace{0.5cm}
\caption{\label{F5}(Color online)
Muon capture transition rate to the $^{12}$B
ground state in units of $10^2s^{-1}$,  and electron and muon folded ECSs to the $^{12}$N
ground state in units of $10^{-42}$ cm$^2$ and  $10^{-41}$ cm$^2$, respectively.
 The experimental
values, in above units, are: $\Lambda(^{12}$B$)=6.2\pm0.3$~\cite{Mil72},
$\overline{\sigma}_e(^{12}$N$)=9.1 \pm 0.4\pm 0.9 $~\cite{Ath97},
and $\overline{\sigma}_e(^{12}$N$)=8.9 \pm 0.3\pm 0.9 $~\cite{Aue01}, and
$\overline{\sigma}_\mu(^{12}$N$)=6.6 \pm 1.0\pm 1.0 $~\cite{Ath97a},
and $\overline{\sigma}_\mu(^{12}$N$)=5.6 \pm 0.8\pm 1.0 $~\cite{Aue02a}.
 }\end{figure}

After establishing  the PQRPA parametrization, we analyze  the
behavior of the ECSs of the ground states in $^{12}$N and $^{12}$B, 
as a function of the size of the configuration space.
Figure~\ref{F2} shows the ECSs for the reaction $^{12}$C($\nu,e^-)^{12}$N
(in units of $10^{-42}$ cm$^2$) for several configuration spaces, and for $t=0$, within
three  different energy intervals.
The top panel represents the DAR region, where experimental
data are available~\cite{Ath97}, and  search for neutrino oscillations
was done~\cite{Ath97,Zei98}.
The middle panel represents the region of interest for supernovae neutrinos, as
pointed out in Refs.~\cite{Aga07,Str09}, while the bottom panel shows the asymptotic
behavior of the cross-section, which becomes almost constant
for $E_\nu\simeq 200$  MeV.
Within the spaces $S_2$ and $S_3$ the calculations  reproduce quite  well the experimental
cross sections in the DAR region, as seen from the first panel.

In Figure~\ref{F3}   we show the calculated ECSs for the reaction $^{12}$C($\nu,e^-)^{12}$N
 within several configuration spaces, but now with different values of the $pp$-coupling.
 From comparison with the experimental data in  the DAR region~\cite{Ath97} one observes  that
 the appropriate values for the coupling $t$ for  s.p. spaces
 $S_4$, and   $S_6$, are, respectively, $t=0.2$, and  $t=0.3$,  i.e., the same as those required
to reproduce the experimental energies and the  GR $B$-values
in   $^{12}$B, and $^{12}$N.
\begin{figure}[t]
\vspace{-1.cm}
\begin{center}
\hspace{-1.cm}
{\includegraphics[width=8.6cm,height=13.5cm]{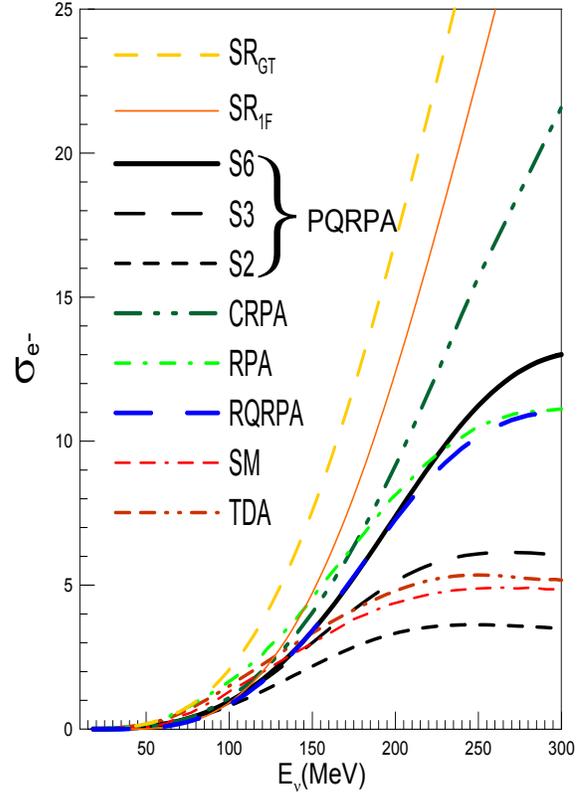}}
\end{center}
\vspace{-1.5cm}
 \caption{\label{F6}(Color online) Inclusive
$^{12}$C($\nu,e^-)^{12}$N cross-section $\sigma_{e^-}(E_\nu)$(in
units of $10^{-39}$ cm$^2$) plotted as a function of the incident
neutrino energy $E_\nu$. The PQRPA results within the s.p. spaces
$S_2$, $S_3$, and $S_6$, and the same values of $s=t$ as in
Figure~\ref{F3}, are compared with two  sum rule limits (as
explained in the text): $SR_{GT}$, and $SR_{1F}$
obtained with average excitation energy
$\overline{\w_{{\sf J}^\pi_n}}$ of   $17.34$, and  $42$ MeV, respectively.
Several previous RPA-like calculations, namely: RPA~\cite{Vol00}, CRPA~\cite{Kol99b}, and
RQRPA within  $S_{20}$ for $E_{2qp}$=100 MeV~\cite{Paa07}, as well as the SM
~\cite{Vol00}, and the TDA~\cite{Con72} are also shown.}
\end{figure}

 This change of parametrization
 hint at the self-consistency of the PQRPA, and
comes from the fact that in this model: i) the GT strength  allocated  in the ground
state  is moved to  another $1^+$ states when the size of the
space is increased, and ii) the effect of the $pp$ residual interaction
goes in the oppositive direction, returning the GT strength to the
$1^+_1$ state. Only for  the space  $S_2$ the cross-section
$\sigma_{e^-}(E_\nu,1^+_{1})$ is appreciable larger (at
$E_\nu\gtrsim 60$ MeV) than for other spaces, which is just
because of the small  number of configurations in this case. In
the same figure are exhibited as well the results for the ECSs
evaluated within the SM~\cite{Eng96}, and the
EPT~\cite{Fuk88}. Both of them agree well with
the data and with the present calculation.

The results for  the reaction
$({\tilde{\nu}},e^+)$ to the ground state in $^{12}$B  are shown in  Figure~\ref{F4}.
The cross-section
 $\sigma_{e^+}(E_{\tilde{\nu}},1^+_{1})$
is similar to that produced by neutrinos but
 significantly smaller in magnitude. When compared with the EPT result
~\cite{Fuk88}, which are also shown in the same figure,
 one notices that they are  considerable
 different.
 To some extent this is surprising as
  in the case of neutrinos the two models yield  very similar results.
One should remember that in the EPT model
   the axial form factor, used for both neutrinos and antineutrinos,
is gauged to the average of the GR $B$-values
in   $^{12}$B, and $^{12}$N, which, in turn, are well reproduced by the PQRPA.
Therefore it is difficult to understand why the EPT results agree with the present calculations
for neutrinos and disagree for antineutrinos.

In Figure~\ref{F5}   we show the dependence on the configuration space
of the exclusive muon capture transition rate $\Lambda(1^+_1)$ to the
$^{12}$B ground state, and the
electron and muon flux-averaged ECSs, given by \rf{2.28}, to the
$^{12}$N ground state, i.e., $\overline{\s}_\e(1^+_1)$, and  $\overline{\s}_\mu(1^+_1)$.
As in Refs.~\cite{Krm02,Krm05} the electron neutrino distribution $n_e(E_\nu)$ was
approximated with the Michel energy spectrum~\cite{Kol99a,Arm02}, and
for the muon neutrinos we used $n_\mu(E_\nu)$ from Ref.~\cite{LSND}. The energy
integration is carried out in the DAR  interval
$m_e+\w_{J_{f}}\le \Delta_{J_{f}}^{\rm DAR}\le 52.8 $ MeV for electrons
and in the  DIF   interval
$m_\mu+\w_{J_{f}}\le \Delta_{J_{f}}^{\rm DIF}\le 300 $ MeV for muons.
From Figure~\ref{F5}, and comparison with experimental data:

  $\Lambda(^{12}$B$)=6.2\pm0.3$~\cite{Mil72},

  $\overline{\sigma}_e(^{12}$N$)=9.1 \pm 0.4\pm 0.9 $~\cite{Ath97},
$8.9 \pm 0.3\pm 0.9 $~\cite{Aue01},

$\overline{\sigma}_\mu(^{12}$N$)=6.6 \pm 1.0\pm 1.0 $~\cite{Ath97a},
$5.6 \pm 0.8\pm 1.0 $~\cite{Aue02a},

\noindent one finds out,   as for results shown in Figures~\ref{F1}, and~\ref{F3}, the model
self-consistency
between s.p. spaces and the $pp$-couplings.  That is, for larger s.p. spaces
 larger values of $t$  are required.
 In brief,   the experimental data of
$\overline{\sigma}_e(^{12}$N), and $\overline{\sigma}_\mu(^{12}$N) are
well reproduced by the  PQRPA.
The same is true for the SM calculations~\cite{Hay00,Vol00}, while in
RPA, and  QRPA models they are strongly overestimated,
as can be seen from Table II in Ref.~\cite{Vol00}, and Table 1 in Ref.~\cite{Sam06}.

\begin{figure}[t]
\vspace{-1.cm}
\begin{center}
\hspace{-1.cm} {\includegraphics[width=9.cm,height=11.cm]{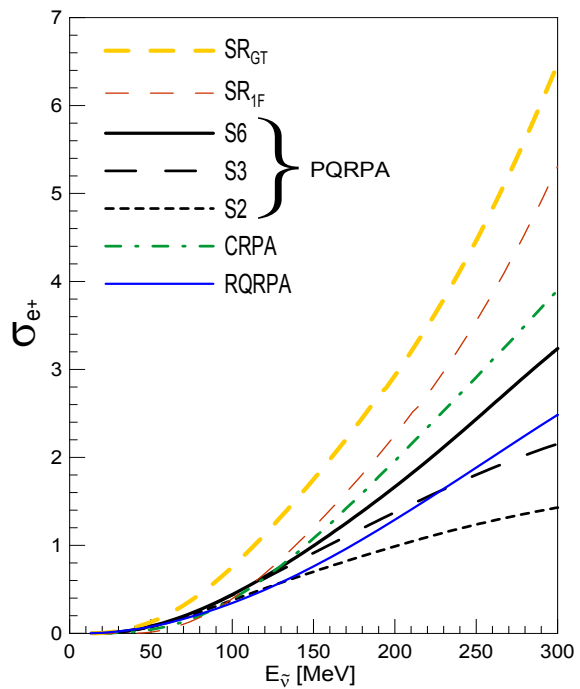}}
\end{center}
\vspace{-1.5cm} \caption{\label{F7}(Color online) Inclusive
$^{12}$C($\tilde{\nu},e^+)^{12}$B cross-section
$\sigma_{e^+}(E_\nu)$ (in units of $10^{-39}$ cm$^2$) plotted as a
function of the incident neutrino energy $E_{\tilde{\nu}}$. All
 results were obtained in the same way as in the
neutrino case in Figure~\ref{F6}.}
\end{figure}
\begin{figure*}[th]
\vspace{-4cm}
\begin{center}
\hspace{-1.cm}
{\includegraphics[width=17.cm,height=18.cm]{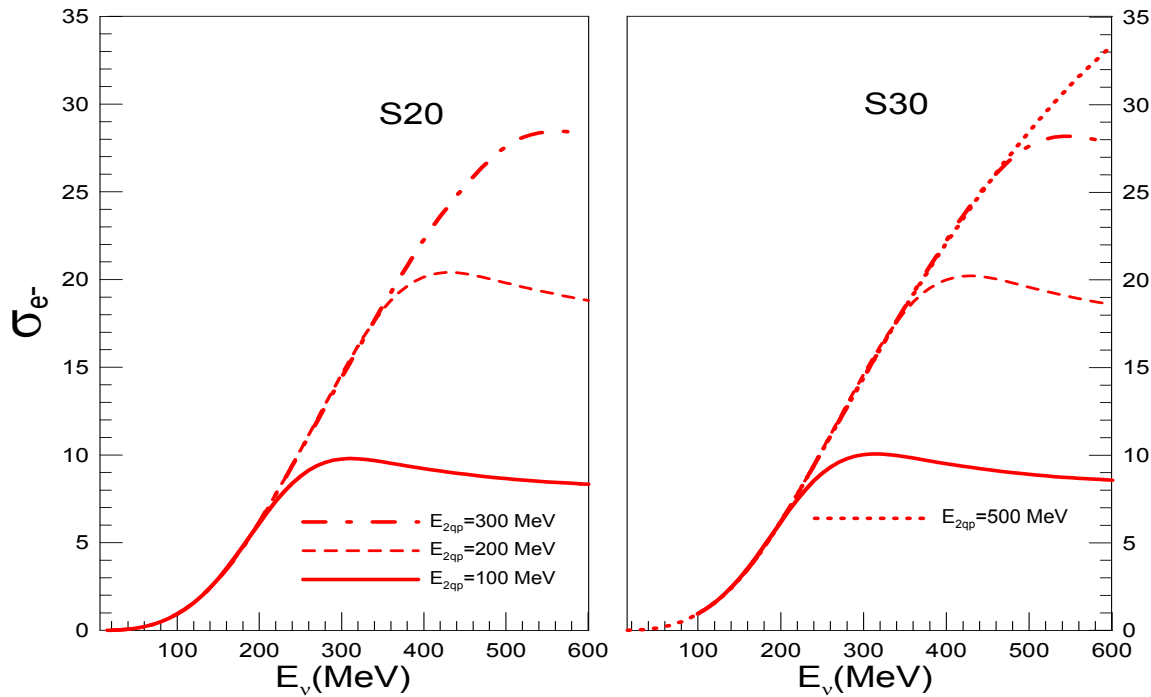}}
\end{center}
\vspace{-5.cm}
 \caption{\label{F8} (Color online) Inclusive
$^{12}$C($\nu,e^-)^{12}$N cross-section $\sigma_{e^-}(E_\nu)$(in
units of $10^{-39}$ cm$^2$) plotted as a function of the incident
neutrino energy $E_\nu$, evaluated in RQRPA with different
configuration spaces. These cross sections are plotted as
functions of the incident neutrino energy with different cut-off
of the $E_{2qp}$ quasiparticle energy as it is explained in the
text. The left and right panels show the cross section evaluated
with $S_{20}$, and $S_{30}$ s.p. spaces. The last cross section
shows that the convergence of the calculation is achieved up to
600 MeV of incident neutrino energy.}
\end{figure*}
\bigskip
\begin{figure*}[t]
\vspace{-4cm}
\begin{center}
\hspace{-1.cm}
{\includegraphics[width=17.cm,height=18.cm]{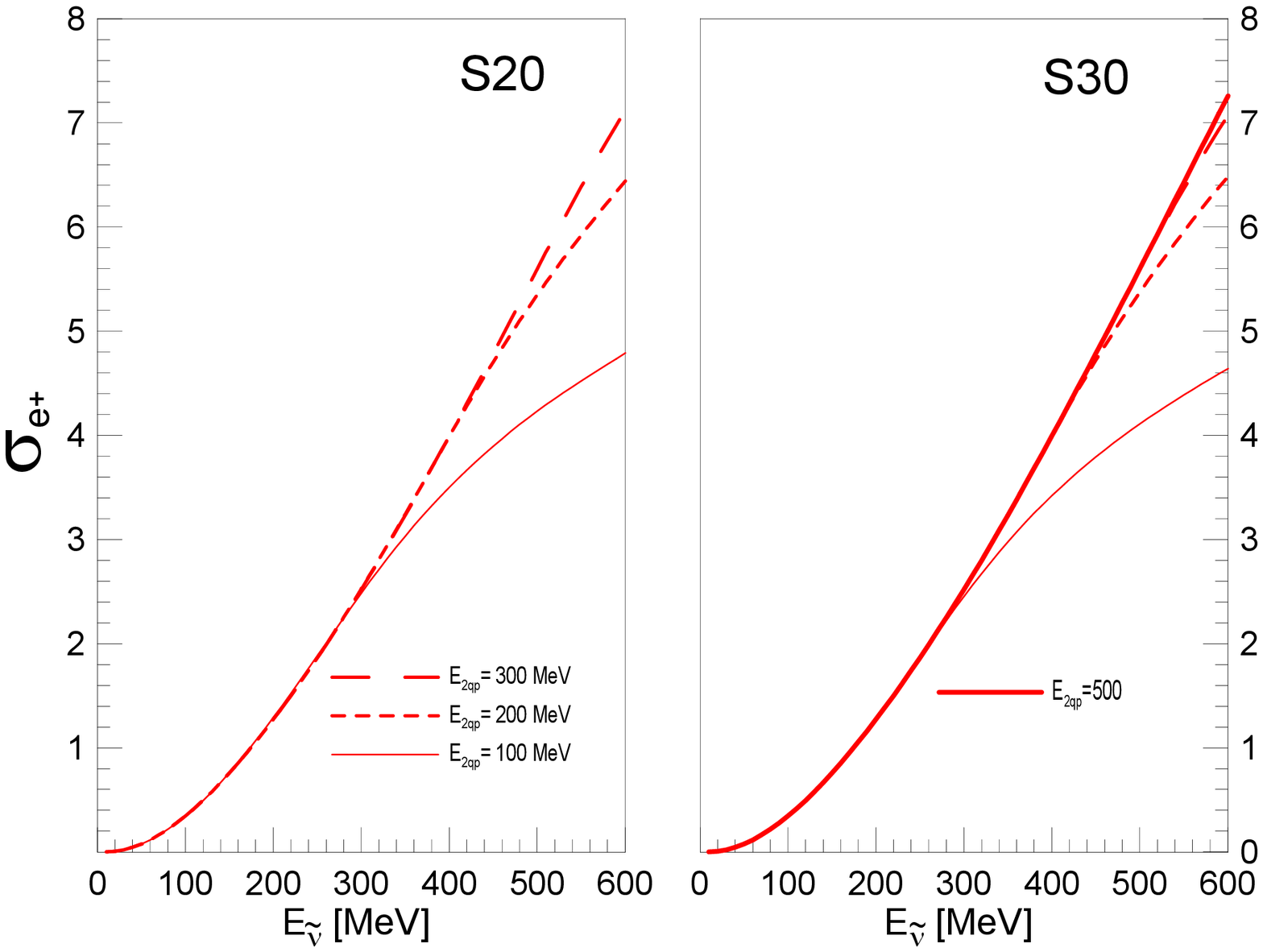}}
\end{center}
\vspace{-5.cm} \caption{\label{F9} (Color online)
Same as in Figure~\ref{F8} but for the
$^{12}$C($\tilde{\nu},e^+)^{12}$B  cross-section
$\sigma_{e^+}(E_{\tilde{\nu}})$.}
\end{figure*}

\subsection{Inclusive  cross-sections $^{12}$C($\nu,e^-)^{12}$N
and $^{12}$C($\tilde{\nu},e^+)^{12}$B, and Sum Rule}
\label{Sec3B}

In Figure~\ref{F6}  we confront the PQRPA results for the ICS
$\sigma_{e^-}(E_\nu)$ within spaces $S_2$, $S_3$, and $S_6$
with the corresponding sum-rules $\s_{e^-}^{SR}(E_{\nu})$
evaluated from \rf{B1}.
One immediately sees that the PQRPA results  depend very strongly
on the size of the employed s.p. space.
On the other hand, as
already mentioned in the Appendix \ref{B}, the sum rule $\s_{e^-}^{SR}(E_{\nu})$
depends on  the average energy  $\overline{\w_{{\sf J}^\pi_n}}$.
Here we use two values
 $\overline{\w_{{\sf J}^\pi_n}}=17.34$ MeV, which is the ground state energy $^{12}$N,
(GT-resonance), and $\overline{\w_{{\sf J}^\pi_n}}=42$ MeV, which is roughly  the energy
of the first forbidden resonance~\cite{Kr83}. The corresponding curves in Figure~\ref{F6}
 are labeled, respectively, as $SR_{GT}$, and $SR_{1F}$.
 They should be the upper limits for allowed and first forbidden
 transitions, respectively.
The validity of these sum rules is questionable for neutrinos energies of
several hundred MeV, as already pointed out by  Kuramoto \etal~\cite{Kur90}.
In fact, we note that the cross section $SR_{GT}$ ($SR_{1F}$)
exceeds the free particle cross section $\s_6\equiv 6\s(\nu_e+n\go e^-+p)$
for $E_\nu>200$ MeV  ($E_\nu>300$ MeV)~\cite{Bud03}.

Several previous SM and RPA-like
calculations of $\s_{e^-}(E_{\nu})$, employing  different effective axial-vector
coupling constants, and
different s.p. spaces, are exhibited in Figure~\ref{F6} as well,
namely:

\noindent
a) TDA~\cite{Con72}, with $g_{\sss A}=1.23$, and  $S_2$,

\noindent b) SM and  RPA~\cite{Vol00}, with $g_{\sss A}=0.88$, and  $S_3$,

\noindent c) CRPA~\cite{Kol99b}, with $g_{\sss A}=1.26$, and $S_{4}$,

\noindent
d) RQRPA  \cite{Paa07},  with $g_{\sss A}=1.23$, $S_{20}$, and $E_{2qp}$=100 MeV.

It is important to specify the values $g_{\sss A}$  because
the partial cross sections are predominantly of the axial-vector type
(specially the allowed ones), which are proportional to $g_{\sss A}^2$.
In spite of very significant  differences in $g_{\sss A}$,
 and the s.p. spaces,
 the different calculations of  $\sigma_{e^-}(E_\nu)$
  yield quite similar results
 for energies $E_\nu\lesssim 130$ MeV, lying
basically  in  vicinity of the
the sum-rule  result $SR_{1F}$.
But for higher energies  they could become quite different, 
and  are clearly separated in two groups at $E_{\nu}=300$ MeV.
In the first group with
$\s_{e^-}(E_{\nu})\lesssim 5\x 10^{-39}$ cm$^2$ are: the SM, TDA, and  PQRPA
within spaces $S_2$, $S_3$, while in the second one with
$\s_{e^-}(E_{\nu})\gtrsim 10\x 10^{-39}$ cm$^2$ are: the RPA, RQRPA, CRPA 
and PQRPA within space $S_6$. Volpe \etal~\cite{Vol00}  have
found that the difference between their SM and RPA calculations
is due to differences in the correlations taken into account, and
 to a too small SM space.  We also note that  only the CRPA result
  approaches the
sum rule limits  for $E_{\nu}> 200$ MeV.

Similar results for  the inclusive
$^{12}$C($\tilde{\nu},e^+)^{12}$B cross-section
 $\sigma_{e^+}(E_{\tilde{\nu}})$
are displayed in Figure~\ref{F7}, and analogous comments can be done here.
For the comparison, we show in the figure the antineutrino-$^{12}$C cross-sections evaluated with
 the CRPA~\cite{Kol99b}.

\subsection{Large configuration spaces}
As there are no experimental data on flux unfolded ICSs for $E_\nu\leq 400$ MeV
 we cannot conclude
which of the results displayed  in Figures~\ref{F6}, and~\ref{F7} are good and which are not.
We can only conclude  that the ICSs strongly depend on the size of the s.p. space.
In the  PQRPA calculations we are not able to use  spaces
 lager than $S_{6}$ because of numerical difficulties.
Thus instead of using the PQRPA, from now on we employ the RQRPA  where such
calculations are feasible. It is important to note that
within the RHB+RQRPA model the oscillator basis is used only in the RHB
calculation in order to
determine the ground state and the single-particle spectra.
The wave functions employed in RPA equations are obtained by
converting the original HO basis to the coordinate representation.
Therefore, the size of the RQRPA configuration
space and $2qp$~ energy cut-offs are determined by the
number of oscillator shells in the RHB model.

 First, we analyze the  effect  of the cut-off energy within  the  $S_{20}$ space on
 $\sigma_{e^-}(E_\nu)$ for $E_\nu$ up to $600$ MeV.
From  the left panel in Figure~\ref{F8} one sees that at  high
energies  this cross-section increases
 roughly  by a factor of two
when $E_{2qp}$ is augmented  from $100$  to $200$ MeV. The
increase of the cross-section is also quite important when
$E_{2qp}$ is moved  from $200$  to $300$ MeV. For the limiting value of $E_{2qp}$=300 MeV, all possible configurations
are included in RQRPA calculations.
   Next, we do the
same within the  $S_{30}$ space,  and the resulting
$\sigma_{e^-}(E_\nu)$ are displayed on the right panel of
Figure~\ref{F8}. From the comparison of both panels it is easy to
figure out that up to $E_{2qp}=300$  MeV the cross sections
obtained with  the $S_{30}$ space are basically the same to those
calculated with the  $S_{20}$ space.
Small differences between the cross sections using $S_{20}$ and  $S_{30}$ spaces for $E_{2qp}$ up
to 300 MeV are caused by modifications of positive-energy single-particle states
contributing to the QRPA configuration space within the restricted $2qp$ energy
window.
But, for $E_\nu\gtrsim 400$ MeV additional transition strength appears
in the $S_{30}$ space when $E_{2qp}$ is moved up to $400$ MeV, from where
further increase in $E_{2qp}$ has  a very small effect. We conclude
therefore that the configuration space engendered by   $N=20$ HO
shells with $E_{2qp}=300$  MeV, is  large enough to describe
$\sigma_{e^-}(E_\nu)$ with $E_\nu$  up to $400$  MeV. Similarly,
the space brought about  by $N=30$ HO shells with $E_{2qp}=400$  MeV
is appropriate
 to account  for $\sigma_{e^-}(E_\nu)$ up to $E_\nu=600$  MeV. For larger neutrino energies
very likely we would have to continue
 increasing  the number of shells.

Analogous results for antineutrino ICSs
$\sigma_{e^+}(E_{\tilde{\nu}})$ are displayed in Figure~\ref{F9}.
One notes important differences in comparison with
$\sigma_{e^-}(E_\nu)$ shown in   Figure~\ref{F8}. First, here the
spaces $S_{20}$ and $S_{30}$ yield almost identical results in the
entire interval of  antineutrino energies up to
$E_{\tilde{\nu}}=600$ MeV. Second, the successive increase
in the cross-sections when the cut-off $E_{2qp}$ is augmented in
steps of $100$  MeV are smaller, and decrease more rapidly    than
in the neutrino case. This suggests that the configuration space
is now sufficiently large to
  appropriately  account for
$\sigma_{e^+}(E_{\tilde{\nu}})$ even at antineutrino energies larger than $600$ MeV.
\begin{figure}[t]
\vspace{-1.cm}
\begin{center}
\hspace{-1.cm} {\includegraphics[width=9.cm,height=11.cm]{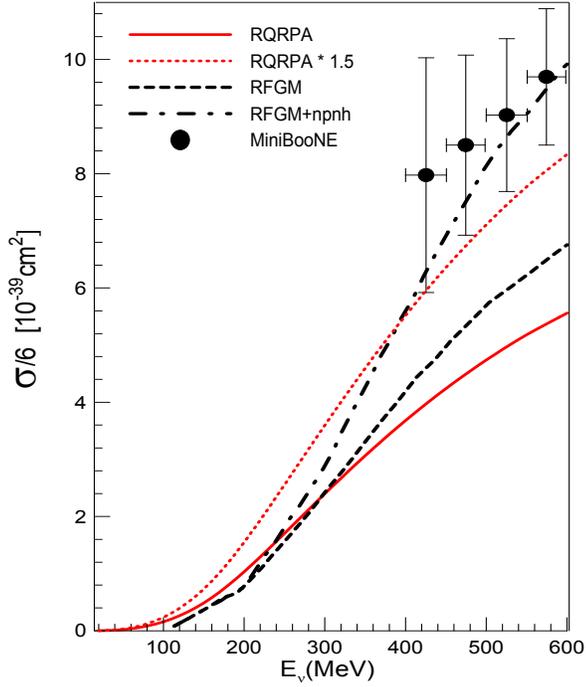}}
\end{center}
\vspace{-1.5cm} \caption{\label{F10}(Color online) The calculated RQRPA
(within $S_{30}$ and $E_{2qp}=500$ MeV) quasi-elastic ($\nu_e,^{12}$C)
cross section per neutron (full line) is compared with that for the
($\nu_\mu,^{12}$C) scattering data measured at MiniBooNE~\cite{MiniBooNE};
with dotted line is shown the same calculation but renormalized by a factor
$1.5$. Also are displayed the
calculations done by
Martini \etal~\cite{Mart09,Mart10} within the RFGM for pure (1p-1h)
excitations (dashed line), and with the inclusion of the np-nh channels
(dot-dashed line).}
\end{figure}

At present, due to numerical difficulties,  we cannot
 perform the RQRPA  calculations for the full range of neutrino energies
 where the QE  cross section was measured at MiniBooNE~\cite{MiniBooNE},
  but only up to 0.6 GeV. However, we feel that it could still be illustrative
for comparison with data.
This is done in Figure~\ref{F10}, which is basically a  piece of  Figure 21 Ref. \cite{Mart09}
for the QE $\sigma_{\mu^-}(E_\nu)$ (see also Ref.~\cite{Mart10}),
where is incorporated our result for $\sigma_{e^-}(E_\nu)$
from Figure~\ref{F8} for $S_{30}$ and
$E_{2qp}=500$  MeV.
As already pointed out in the Introduction,
at relatively  high energies ($E_\nu>300$  MeV) the electron and muon neutrino cross
sections converge to each other, and  therefore, in the present
analysis, the electron neutrino cross section
provides a reasonable upper limit estimate.
 One sees that we
underestimate the  data  by almost a factor of two. But
one should keep in mind
that, while we use  $g_{\sss A}=1$ (see \rf{2.4}) in the RFGM calculation done
by Martini \etal~\cite{Mart09,Mart10}  $g_{\sss A}=1.255$
was used. Being the axial-vector contribution dominant for the latter value
of the coupling constant, one would have to  re-normalize
our $\sigma_{e^-}(E_\nu)$ by a factor $\sim 1.5$. Such a result is also shown
in Figure ~\ref{F10}, and although the resulting cross section  still underestimates somewhat
the data for $\sigma_{\mu^-}(E_\nu)$, it is notably superior to the
  pure 1p-1h  result from Ref.~\cite{Mart09}, where good agreement with the data
is achieved only after considering additional two-body (2p-2h) and three-body (3p3h)
decay channels. One should keep in mind, however, that
as the weak decay Hamiltonian  is   one-body
operator, these  excitations are only feasible via the
ground-state correlations (GSC), which basically redistribute the
1p-1h transition strength without increasing its total magnitude when
the  initial wave function is properly normalized. In the present
work, as well as in all SM-like calculations, the GSC, and a
normalized initial state wave function are certainly considered
to all orders in perturbation theory  through  the full
diagonalization of the hamiltonian matrix.
On the other hand, in Refs.~\cite{Mart09,Mart10} the GSC are taken  into account
 in second order perturbation theory, but there are no references
to the normalization of the $^{12}$C ground-state wave function.
How to carry out the normalization in the framework of the
infinite nuclear matter model  is discussed in a recent paper
related to the nonmesonic weak decay of the hypernucleus
$^{12}_\Lambda$C~\cite{Bau10}; see also Refs.~\cite{Ma91,Va92,Ma95}.

\subsection{Multipole decomposition of cross-sections}

\begin{figure*}[th]
\vspace{4cm}
\begin{center}
 {\includegraphics[width=17.2cm,height=10.cm]{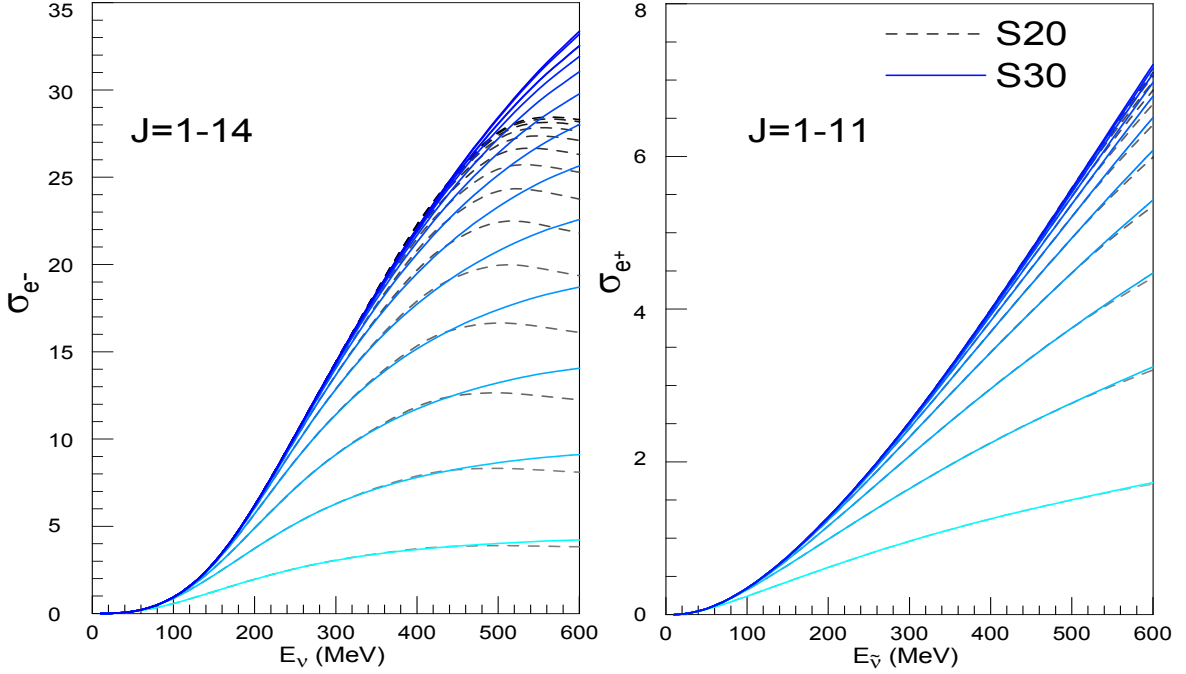}}
\end{center}
\vspace{-5.cm}
\caption{\label{F11} (Color online) Left and right panels
show, respectively,   the cross sections $\sigma_{e^-}(E_\nu)$, and
$\sigma_{e^+}(E_{\tilde{\nu}})$ (in units of $10^{-39}$ cm$^2$)
evaluated in RQRPA for  $S_{20}$, and  $S_{30}$ s.p. spaces with the cutoff
 $E_{2qp}=500$ MeV, and  different
maximal values of ${\sf J}^\pm$, with ${\sf J}$ going from $1$ up
to $14$ for neutrinos, and from $1$ up
to $11$ for antineutrinos.}
\end{figure*}

We did not mention yet the contributions of
different multipoles to the ICSs. Normally, the RHB model within   $S_{20}$, and with ${\sf
J}^{\pi}\le 7^{\pm}$, provides converged results for RQRPA
excitation spectra at incident neutrino energies $ E_\nu\leq 300$
MeV as seen from Figure 2 in Ref.~\cite{Paa07}. But this is not the case for neutrino-nucleus
cross sections at  energies $E_\nu\gtrsim 300$ MeV where one has to consider large cutoff energies
$E_{2qp}$. In fact, it is necessary to consider more and more
 multipoles according as the configuration space is enlarged by increasing
$E_{2qp}$. This is illustrated in Figure~\ref{F11} for the case of $E_{2qp}=500$ MeV.
One sees that are significant all multipoles up to  ${\sf
J}^{\pi}=14^{\pm}$ for neutrinos,  and up to  ${\sf
J}^{\pi}=11^{\pm}$ for antineutrinos.

\begin{figure*}[t]
\begin{center}
\begin{tabular}{cc}
\hspace{-.5cm}
{\includegraphics[width=9.cm,height=11.cm]{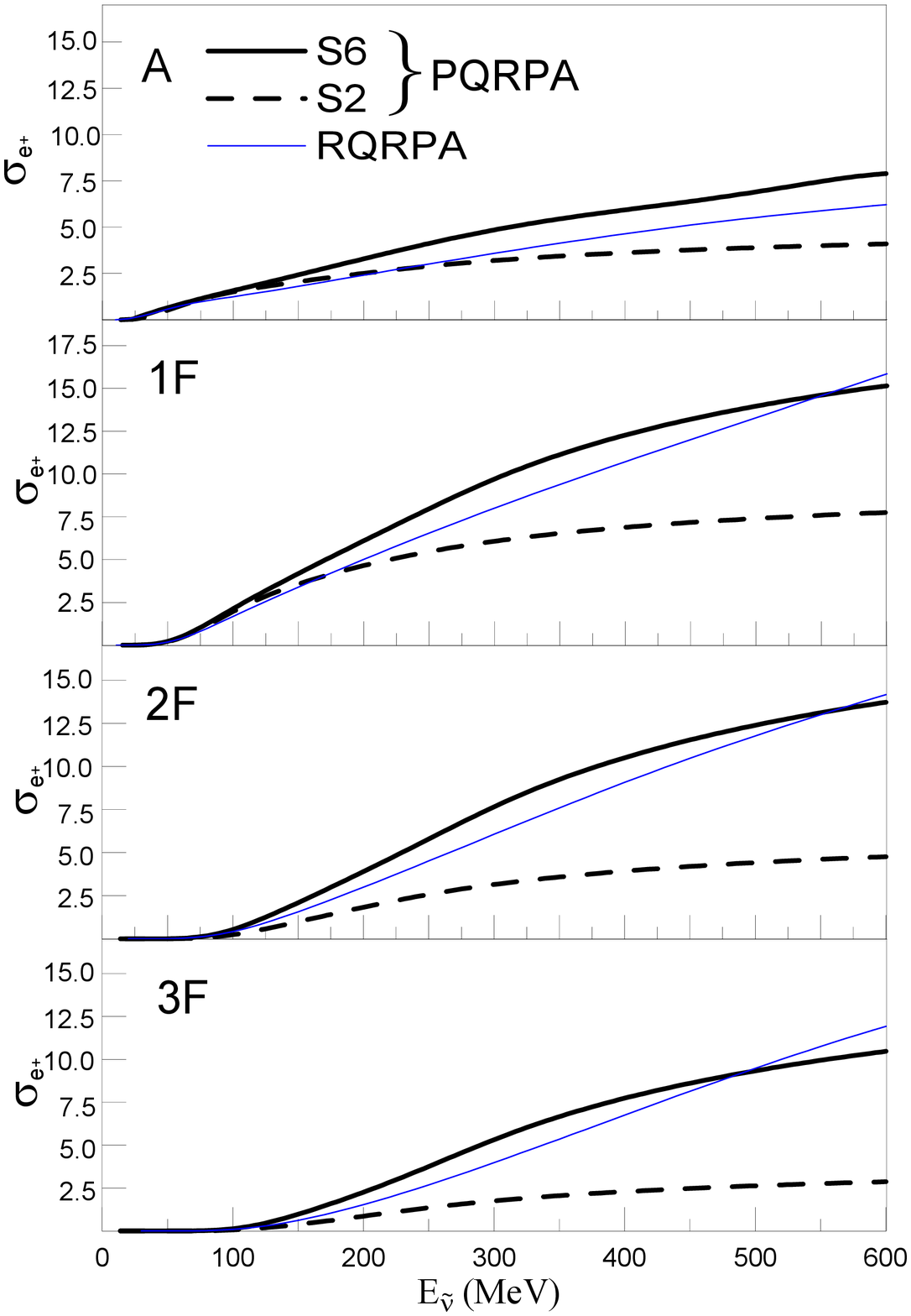}}
&\hspace{-1.cm}
{\includegraphics[width=9.cm,height=11.cm]{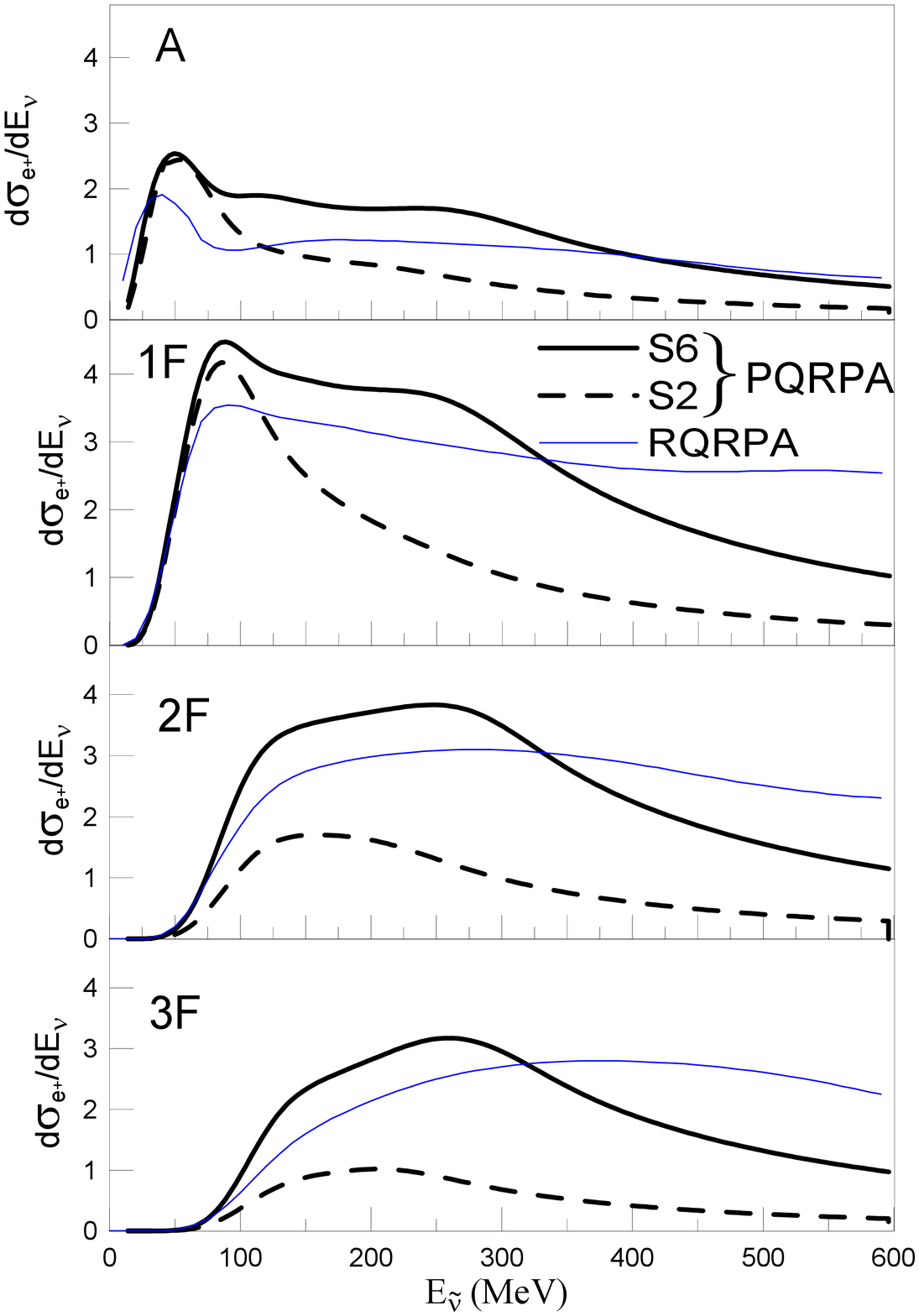}}
\end{tabular}
\end{center}
\vspace{-1cm} \caption{\label{F12}(Color online) Left panel:
Allowed ($J^\pi=0^+,1^+$), first-forbidden ($J^\pi=0^-,1^-,2^-$),
second-forbidden ($J^\pi=2^+,3^+$), and third-forbidden
($J^\pi=3^-,4^-$) inclusive $^{12}$C($\tilde{\nu},e^+)^{12}$B
cross-section $\sigma_{e^+}(E_{\tilde{\nu}})$ (in units of
$10^{-42}$ cm$^2$), plotted as a function of the incident neutrino
energy $E_{\tilde{\nu}}$. Right panel: Same as left panel but now
for $d\sigma_{e^+}(E_{\tilde{\nu}})/dE_{\tilde{\nu}}$ (in units of
$10^{-42}$ cm$^2$ MeV$^{-1})$.
}
\end{figure*}

Next we discuss the partial multipole
contributions to the ICS, having in view the degree of   forbiddenness of the
transition matrix elements, namely,

$\bullet$ Allowed: $\sigma^A_{e^+}(E_{\tilde{\nu}})$,  with ${\sf J}^{\pi}=0^+,1^+$,

$\bullet$ First-forbidden $\sigma^{1F}_{e^+}(E_{\tilde{\nu}})$, with  ${\sf J}^{\pi}=0^-,1^-,2^-$,

$\bullet$ Second-forbidden $\sigma^{2F}_{e^+}(E_{\tilde{\nu}})$,  with ${\sf J}^{\pi}=2^+,3^+$, and

 $\bullet$ Third-forbidden $\sigma^{3F}_{e^+}(E_{\tilde{\nu}})$ with ${\sf J}^{\pi}=3^-,4^-$,

 \noindent cross-sections.
  Thus, in the left panel of
Figure~\ref{F12}  we show these individual contributions for the inclusive
$^{12}$C($\tilde{\nu},e^+)^{12}$B cross-section $\sigma_{e^+}(E_{\tilde{\nu}})$,
  evaluated within both the PQRPA (spaces $S_2$, and $S_6$) the RQRPA
(space $S_{30}$ with  $E_{2qp}=500$ MeV).

 The same is done for the corresponding  derivatives, i.e., the spectral
functions $d\sigma_{e^+}(E_{\tilde{\nu}})/dE_{\tilde{\nu}}$,
on the right panel of the same figure.  Several conclusions can be drawn.
First, as in the case of  total
$\sigma_{e^+}(E_{\tilde{\nu}})$, they depend very strongly on the size of the
configuration space. This dependence, in turn, increases with the degree of
forbiddenness; that is,
it is more pronounced for  first-forbidden than for allowed transitions, and so on.
Second, within the PQRPA the allowed cross-section $\sigma^A_{e^+}(E_{\tilde{\nu}})$
 exhibits a resonant pattern at low energy, and
is   dominant for $E_{\tilde{\nu}}\lesssim 50$ MeV.
For large s.p. spaces its contribution is quite significant even at $E_{\tilde{\nu}} =500$
MeV.
\footnote{The denominations   here don't have exactly the same meaning as in the low-energy $\beta$-decay,
where allowed transitions are those within the same HO shell ($\Delta N=0$), while here
 all values of $\Delta N$ are permitted. Similarly happens with the forbidden transitions.
 The degrees of hindrance basically come from value of the orbital angular momenta.     }
In the case of RQRPA, the spectral function $d\sigma^A_{e^+}(E_{\tilde{\nu}})/dE_{\tilde{\nu}}$ also displays
low-energy resonant structure, and $\sigma^A_{e^+}(E_{\tilde{\nu}})$ is always smaller
in magnitude than in the PQRPA case.
  Third, $\sigma^{1F}_{e^+}(E_{\tilde{\nu}})$ is   peaked at $E_{\tilde{\nu}}\sim 75$ MeV, and its
contribution is always larger than that of $\sigma^A_{e^+}(E_{\tilde{\nu}})$ for
$E_{\tilde{\nu}}\gtrsim 150$ MeV.
Fourth, $\sigma^{2F}_{e^+}(E_{\tilde{\nu}})$, and  $\sigma^{3F}_{e^+}(E_{\tilde{\nu}})$
mainly contribute in the interval $150\lesssim E_{\tilde{\nu}}\lesssim 400$ MeV, and
their overall contributions are of the same order of magnitude, and comparable to that of the
$\sigma^{1F}_{e^+}(E_{\tilde{\nu}})$. Fifth, the contributions of the remaining  multipoles with
${\sf J}^{\pi}=4^+,5^\pm,6^\pm,7^\pm$  are always  very small for the space $S_2$,
but are quite sizeable
for  $S_6$ at high energies. For instance, at
$E_{\tilde{\nu}}=100,300,600$ MeV they contribute, respectively with  $0.02\%,0.86\%,1.18\%$
within the space $S_2$,
and  $0.04\%,14\%,20\%$ for $S_6$.
With further increase of the single-particle basis, configurations from higher
multipoles become
more pronounced at higher neutrino energies. In particular, the sum of contributions
coming from
${\sf J}^{\pi}=4^+,\cdots, 11^\pm$
multipoles when evaluated within the  RQRPA using the space $S_{30}$ and maximal
 value of $E_{2qp}$=500 MeV, are 1.1\%, 14.4\%, and 33.2\%
 at $E_{\nu}$=100, 300, and
600 MeV, respectively.

Recently Lazauskas and Volpe~\cite{Laz07,Vol04} have suggested the convenience
of performing nuclear structure studies using low energy
neutrino and antineutrino beams. Because of feasibility reasons the flux
covers $80$ MeV only.
Nevertheless, from the analysis of $^{16}$O, $^{56}$Fe,
 $^{100}$Mo and $^{208}$Pb nuclei within the QRPA using the Skyrme force
 they were able to disentangle  the  multipole distributions of  forbidden cross-sections,
showing that the forbidden multipole contribution is different for various nuclei.
In this work  we extend this kind of study to $^{12}$C.

 In Table \ref{table} we show the results for the
 flux-averaged cross sections $\overline{\s}_{e^+}$ for the reaction
$^{12}{\rm C}({\tilde{\nu}},e^+)^{12}{\rm B}$. In \rf{2.29} we have used the same
antineutrino fluxes $n_{e^+}(E_{\tilde{\nu}})$ as in Ref.~\cite{Laz07},  i.e.,
the DAR flux, and those produced by boosted $^6$He ions with different
values of time dilation factor $\gamma=1/\sqrt{1-v^2/c^2}$.
 Results of two calculations are presented: i) PQRPA within  $S_6$,
and ii) RQRPA  within $N=20$, and cutoff  $E_{2qp}=300$ MeV. One
sees that in both models, and principally in the PQRPA, the
allowed transitions dominate the forbidden one, and specially for
the low-energy beam with $\gamma=6$. The contributions of the
second-forbidden processes are very small in all the cases, while
those coming from third-forbidden ones are always negligible.
All this is totally consistent with the results shown in Figure~\ref{F12},
from where it is clear that to study second and third forbidden
reactions in $^{12}$C, one would need fluxes $n_{e^+}(E_{\tilde{\nu}})$ with
$E_{\tilde{\nu}}$ at least up to  $\gtrsim 150$ MeV.
It should also be stressed that our results both for allowed and
forbidden transitions fully agree with those obtained in
Ref.~\cite{Laz07}; the difference in $^{16}$O comes from the
double-shell closure in this nucleus.

\begin{table}[ht]
\caption{\label{table}~Fraction (in \%) of  flux-averaged cross
sections $\overline{\s}_{e^+}$ for $^{12}{\rm C}({\tilde
\nu},e^+)^{12}{\rm B}$ for allowed (A), first-forbidden (1F),
second-forbidden (2F), and third-forbidden (3F) processes. The
antineutrino fluxes $n_{e^+}(E_{\tilde{\nu}})$  are the same as in
Ref.~\cite{Laz07}, i.e., the DAR flux, and those produced by
boosted $^6$He ions with different values of
$\gamma=1/\sqrt{1-v^2/c^2}$.
 Results of two calculations
 are presented: i) PQRPA within  $S_5$,
and ii) RQRPA  within $N=30$, and cutoff  $E_{2qp}=300$ MeV.
}%
\begin{ruledtabular}
\newcommand{\cc}[1]{\multicolumn{1}{c}{#1}}
\begin{tabular}{ c | c | c c c}
&DAR&\multicolumn{3}{c}{$\gamma$}\\
&     &6&10&14\\ \hline
A&&&&\\
PQRPA&79.43&92.09&77.00&63.01\\
RQRPA&84.40&94.88&82.25&67.15\\
1F&&&&\\
PQRPA&20.03& 7.83&22.16&33.76\\
RQRPA&15.10& 4.13&16.86&29.61\\
2F&&&&\\
PQRPA& 0.51& 0.07& 0.78&2.89 \\
RQRPA& 0.55& 0.08& 0.81&2.91 \\
3F&&&&\\
PQRPA& 0.018&0.002&0.04&0.33 \\
RQRPA& 0.025&0.011&0.05&0.33 \\
\end{tabular}
\end{ruledtabular}
\end{table}

\subsection{Supernova neutrinos}
\begin{figure}[th]
\begin{center}
{\includegraphics[width=8.6cm,height=10.cm]{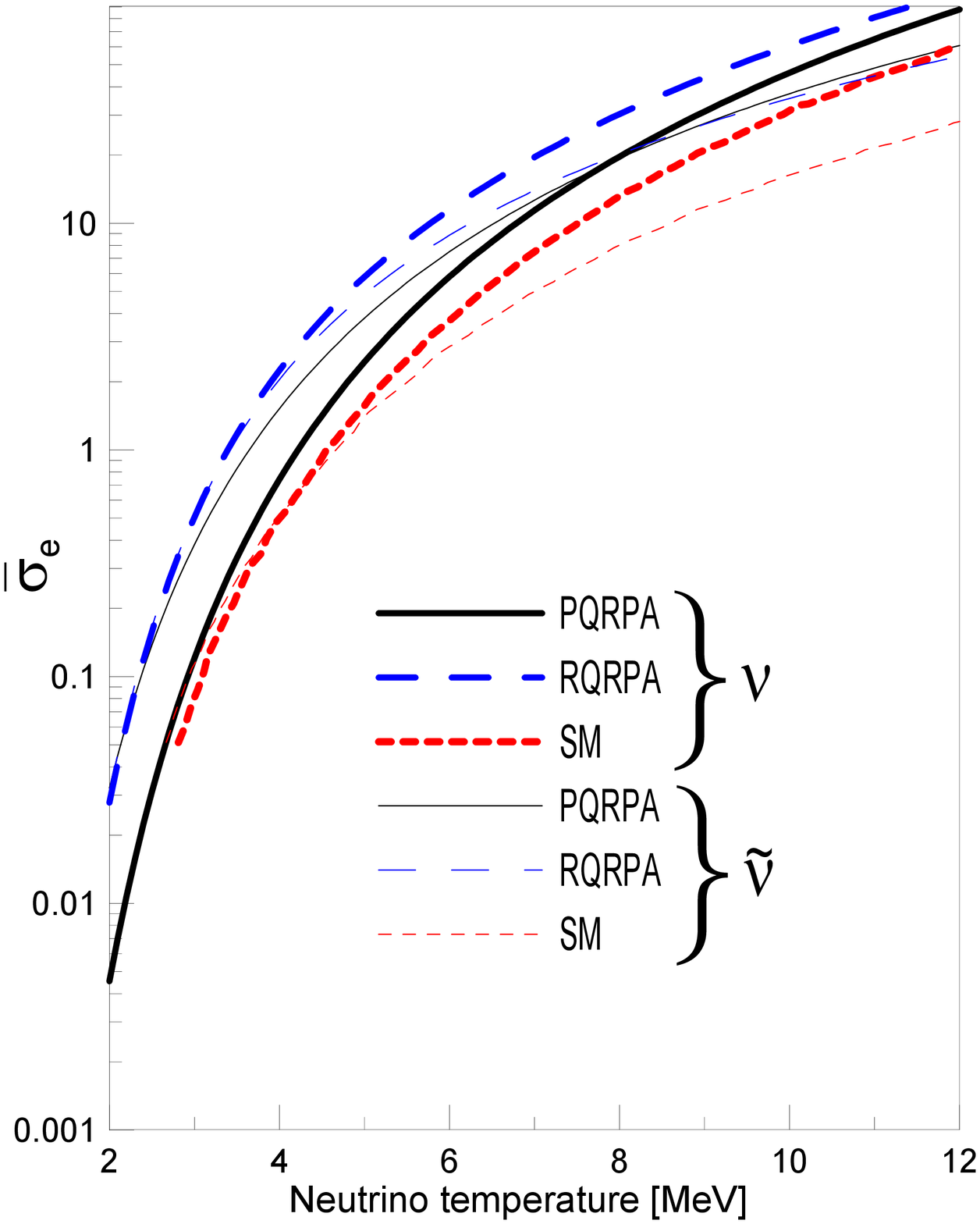}}
\end{center}
\vspace{-1.cm} \caption{\label{F13} (Color online)
Flux-averaged neutrino and antineutrino cross
sections $\overline{\s}_{e^\pm}$ in $^{12}$C
with typical supernovae fluxes.
}%
\end{figure}

We also  address briefly the $\nu/\tilde{\nu}$-$^{12}$C nucleus
cross-sections  related with astrophysical applications,
the knowledge of which could  have important implications.
For this  purpose,
 are evaluated the  $\overline{\s}_{e^\pm}$ folded with
  supernovae $\nu/\tilde{\nu}$ spectra represented by a normalized
Fermi-Dirac distribution with temperatures in the interval $T_{\nu_e}=2-12$ MeV, which
includes the most commonly used values
are $T_{\nu_e}=3.2$ MeV and $T_{\bar{\nu}_e}=5.0$ MeV.
 For   mean
energies  $\langle E_\nu \rangle \approx 3.15 \times T_\nu$, and
zero chemical potential \cite{Woo90,Kei03} the neutrino  flux is
\be
n_e(E_\nu)=\frac{0.5546}{T_\nu^3}
\frac{E_\nu^2}{e^{E_\nu/T_\nu}+1},
 \label{3.5}\ee
and similarly for antineutrinos. For the sake of
simplicity we do not analyze same   relevant aspects of
$n_e(E_\nu)$ in supernova simulation, such as the MSW effect (see,
for example, Ref.~\cite{Akh00}), and the spectral swapping of the
neutrino flux (Ref.~\cite{Dun07}). In Figure \ref{F13}   we confront the
 $\nu-^{12}$C cross sections averaged over
supernova $\nu$-fluxes for the range of  $T_\nu=2-12$ MeV, obtained within following calculations:

\noindent
i) PQRPA within $S_6$,

\noindent ii) RQRPA within $S_{30}$ and  $E_{2qp}=500$ MeV, and

\noindent iii) SM done by Suzuki \etal~\cite{Suz06} with the SFO Hamiltonian (the PSDMK2
interaction yields a quite similar result).

\noindent As seen from Figure \ref{F13}, in the vicinity of the temperatures mentioned at the beginning
($T_\nu=3-5$ MeV), these three calculations yield, respectively, that:
i) $\overline{\s}_{e^+}$ is significantly larger than
$\overline{\s}_{e^-}$,  ii)
$\overline{\s}_{e^-}$ is only slightly larger than $\overline{\s}_{e^+}$, and
iii)  $\overline{\s}_{e^+}\cong\overline{\s}_{e^-}$.
Both SM  cross sections are always smaller than those obtained in the
the other two calculations, and specially in comparison with
the RQRPA one.

\section{Summary and concluding remarks}\label{Sec4}

The present work is a continuation of our previous works~\cite{Krm02,Krm05}.
In fact, the formalism for weak interaction processes introduced there
is now elaborated more thoroughly yielding very  simple expressions
for the transition rates,  which  greatly facilitate the numerical calculation.
This is done through the separation
of the nuclear matrix elements into their real and imaginary
parts, which, in turn,  permits to split  the transition rates,
 for neutrino-nucleus reactions \rf{2.22}
 into natural \rf{A6} and unnatural
 parity  \rf{A7} operators. Similar separation is done for  the  muon capture transition rate
 \rf{D3} in Eqs.  \rf{D4} and  \rf{D5}.
 Moreover, consequences of explicit violation of CVC hypothesis by the Coulomb field \rf{2.18}
are addressed for the first time, and the sum rule approach for the inclusive cross section,
proper  to the present formalism,
has been worked out in the Appendix B. For the sake of completeness, the extreme
relativistic limit of neutrino-nucleus cross section  is also presented in the Appendix C,
where in the formula for transition rates turn out to be still simpler. We note that, except at very
low neutrino energies, they can be used without any restriction in practical applications.

We have  discussed in details the inclusive properties that comprise:

i) Ground state
energies in  $^{12}$B and $^{12}$N, and the corresponding GT $B$-values (Figure ~\ref{F1}),

ii) Exclusive $^{12}$C($\nu,e^-)^{12}$N cross-section
$\sigma_e(E_\nu,1^+_{1})$, as a function of the incident neutrino
energy $E_\nu$ (Figures~\ref{F2}, and ~\ref{F3}),

iii) Exclusive $^{12}$C(${\tilde{\nu}},e^+)^{12}$B cross-section
$\sigma_{e^+}(E_{\tilde{\nu}},1^+_{1})$, as a function of the incident antineutrino
energy $E_{\tilde{\nu}}$ (Figure ~\ref{F4}), and

iv)  Muon capture transition rate to the $^{12}$B
ground state,  and electron and muon folded cross-sections to the $^{12}$N
ground state $\overline{\s}_\e(1^+_1)$, and  $\overline{\s}_\mu(1^+_1)$ (Figure ~\ref{F5}).

\noindent Special attention was paid to the interplay between
 the size of the configuration space,
 and the magnitude of the residual interaction within the
$pp$-channel. It was found that as the first becomes larger, the second has to increase
to obtain the agreement with the experimental data for the exclusive observables.

The main purpose of our discussion of exclusive
properties was to put in evidence  the limitations
of the RPA and the QRPA models.
The basic problem in the implementation of  the  RPA is
the lack of pairing correlations, i.e.
the inability for opening the $1p_{3/2}$ shell, while deficiency
of the standard QRPA is in the non-conservation of the number
of particles, as evidenced by the wave
functions \rf{3.1}, \rf{3.2}, and \rf{3.3} presented in Sec.~\ref{Sec3}.
In this way  we have definitively established that
the SM and  the PQRPA are
proper theoretical frameworks to describe the
ground state properties of $^{12}$B and  $^{12}$N.
\footnote{ After  our work has been  finished,
 Cheoun \etal~\cite{Ch10} have presented a new  evaluation of the
 ECS in $^{12}$C within the QRPA. They  get  good agreement with data
 for $\overline{\sigma}_e(^{12}$N), which is at variance
with the previous QRPA calculation~\cite{Vol00}.}

The inclusive  cross-sections $^{12}$C($\nu,e^-)^{12}$N
and $^{12}$C($\tilde{\nu},e^+)^{12}$B have been studied within
the PQRPA in the same manner as the exclusive  ones for $E_{\nu_e}$ up to $300$ MeV.
As there are  no experimental available
in this case the  comparison is done with the previous calculations  only
\footnote{
As already mentioned in the introduction the only available
experimental data on $^{12}$C ICS  is
the low neutrino energy  ($E_{\nu_e}<60$ MeV) folded one,
which has already been discussed in our previous works \cite{Krm05,Sam06,Paa07}.},
and displayed  in Figures \ref{F6}, and \ref{F7}. Here, unlike within
Figures \ref{F2}, \ref{F3}, and \ref{F4},
 we also show the results obtained with the other RPA-like
 models ~\cite{Con72,Vol00,Kol99b,Paa07},
which could  be  a suitable framework for  describing   global nuclear properties
such as it is the inclusive  cross-sections.

When the  size of the configuration space is enlarged the  calculated PQRPA cross-sections,
 at difference with the exclusive ones, steadily increase, and particulary
 for neutrino energies larger than $200$ MeV,
 in spite of including the particle-particle interaction.  At low energy
 they approach to  the cross section
 of the first-forbidden sum-rule limit, but
 are significantly smaller at high energies both for neutrino and antineutrino.

The largest space that we can deal within  the number projection procedure
is the one that includes all the orbitals
until the $N=6$ HO shell. This is the reason why we have recurred to the
RQRPA where it is possible to employ larger configuration spaces. It seems that
when the number of shells is increased to $N={30}$, and the  cut-off energy
$E_{2qp}$ is large enough, the cross sections very likely converge
as shown in Figures \ref{F8}, \ref{F9},  \ref{F10}, and  \ref{F11}.

The Figure
 \ref{F10} also indicates  that  the RQRPA is a promising nuclear model to reproduce
  the quasi elastic ($\nu_\mu,^{12}$C) cross section in the region of $E_{\nu_\mu}\sim 1$ GeV
 which has been measured recently at MiniBooNE~\cite{MiniBooNE}. We do not know whether the
 discrepancy between the experiment and the theory comes from the
 non completeness of the configuration space or from the smallness
of the effective axial-vector coupling constant that we are using $\gA=1$. It could also happens
that we need $\gA=1$ for the low energy exclusive cross section
and $\gA=1.255$ for the high energy inclusive
cross section. We do not understand the reason for such a energy dependence of $\gA$, but it is consistent with
the Eq. (23) in Ref.~\cite{Na82} where it is shown that  for the low energy $\beta$-decay $\gA$ could be
much more quenched that the total GT strength. We hope to be able to say more on this matter in the
next future.

We have also addressed the issue of multipole composition of the inclusive cross sections,
by separating them into allowed
($J^\pi=0^+,1^+$), first-forbidden ($J^\pi=0^-,1^-,2^-$),
second-forbidden ($J^\pi=2^+,3^+$), and third-forbidden
($J^\pi=3^-,4^-$) processes. The results for  the antineutrino reaction $^{12}$C($\tilde{\nu},e^+)^{12}$B
are displayed in Figure~\ref{F12} both for the PQRPA and the RQRPA. Of course,
 similar  results are obtained also for neutrinos.
 We remark  that the spectral functions $d\sigma^A_{e^+}(E_{\tilde{\nu}})/dE_{\tilde{\nu}}$,
  when evaluated within the PQRPA,
clearly put into evidence the resonant structure of the allowed cross-section,
 which is mainly of the GT type.

The study of the partial ICSs has been related with the proposal
done in Refs.~\cite{Laz07,Vol04}
of performing nuclear structure studies of forbidden processes by using low energy
neutrino and antineutrino beams. From the results shown in Table \ref{table}
 for the flux-averaged cross sections $\overline{\s}_{e^+}$ in the reaction
$^{12}{\rm C}({\tilde{\nu}},e^+)^{12}{\rm B}$ we show  that the contribution
of allowed transitions decreases gradually in favor of the
first forbidden transitions according with the increase of $\gamma$-boost.
We conclude  that to study high forbidden
reactions one would need  ${\tilde{\nu}}$-fluxes with
$E_{\tilde{\nu}}$ up to  $\gtrsim 150$ MeV in $^{12}C$.

At the end we considered possible astrophysical applications of
the $\nu/\tilde{\nu}$-$^{12}$C nucleus
folded  cross sections $\overline{\s}_{e^\pm}$, using
supernovae $\nu/\tilde{\nu}$ spectra represented by a normalized
Fermi-Dirac distribution with mean
energy $\langle E_\nu \rangle \approx 3.15 \times T_\nu$, and
zero chemical potential.  It is found that for temperature $T_\nu=3-5$ MeV
both the PQRPA and  RQRPA models yield significantly larger  cross sections
that the previously used shell model.

\section*{Acknowledgements}
This work was partially supported by the Argentinean agency CONICET under
contract PIP 0377, and by the U.S. DOE grants
DE-FG02-08ER41533, DE-FC02-07ER41457 (UNEDF, SciDAC-2) and the Research Corporation.
A.R.S. thanks to W.C. Haxton and G.M. Fuller for stimulating
discussion and to the Institute of Nuclear Theory
of University of Washington, where part of this work was performed.
N. P. acknowledges support by the Unity through Knowledge Fund
(UKF Grant No. 17/08),  MZOS - project 1191005-1010
 and Croatian National Foundation for Science.

\begin{appendix}
\section{Contributions to $\T_{{\sf J}^\pi_n}(\k)$ of
 natural and unnatural parity states\label{A}}
The real and imaginary parts of the operators ${\sf O}_{{\a}{\sf
J}}$ given by \rf{2.12} and \rf{2.20} do not contribute
simultaneously. In fact,  the  $\Re {\sf O}_{{\a}{\sf J}}$ ($\Im
{\sf O}_{{\a}{\sf J}}$) contributes to  natural (unnatural) parity
states, which
  means that we always can work only with real operators, which greatly simplifies the calculations.
To see this we note that, while  the operators  $\M^{\sss V}_{\sf
J}$, $\M^{\sss A}_{\sf J}$, and \br {\M}^{\sss A}_{0{\sf J}}
&=&\sum_{{\sf L}= {\sf J }\pm1}(-)^{({\sf J-L}-1)/2}\ F_{{{\sf
LJ}0}}j_{\sf L}(\rho) \left[Y_{{\sf
L}}(\hat{\rb})\otimes{\mbs}\right]_{{\sf J}},
\nn\\
 \label{A1}
\er
 are real,  $\M^{\sss A}_{\pm1{\sf J}}$ and $\M^{\sss V}_{\pm1{\sf J}}$
are not. Explicitly,
\br \M_{\pm1{\sf J}}^{\sss A}&=&\M_{\pm1\sf J}^{\sss
A,R}+i\M_{\pm1\sf J}^{\sss A,I}
\nn\\
\M_{\pm1{\sf J}}^{\sss V}&=&\M_{\pm1\sf J}^{\sss V,R}+i\M_{\pm1\sf
J}^{\sss V,I} \label{A2}\er where
 \br
  \M^{\sss A,R}_{ 1\sf J}&\equiv&\M^{\sss A,R}_{ -1\sf J}
 \nn\\
 &=&\sum_{{\sf L}={\sf J }\pm1}(-)^{({\sf J-L}-1)/2}\ F_{ {{\sf LJ}1}}j_{\sf L}(\rho) \left[Y_{{\sf
L}}(\hat{\rb})\otimes{\mbs}\right]_{{\sf J}},
\nn\\
\M^{\sss A,I}_{1{\sf J}}&\equiv&-\M^{\sss A,I}_{-1{\sf J}}=-F_{
1{\sf JJ}}j_{\sf J}(\rho) \left[Y_{{\sf
J}}(\hat{\rb})\otimes{\mbs}\right]_{{\sf J}},
 \nn\\
  \M^{\sss V,R}_{ 1\sf J}&\equiv&\M^{\sss V,R}_{ -1\sf J}
 \nn\\
 &=&\sum_{{\sf L}={\sf J }\pm1}(-)^{({\sf J-L}-1)/2}\ F_{ {{\sf LJ}1}}j_{\sf L}(\rho) \left[Y_{{\sf
L}}(\hat{\rb})\otimes{\mbn}\right]_{{\sf J}},
\nn\\
\M^{\sss V,I}_{1{\sf J}}&\equiv&-\M^{\sss V,I}_{-1{\sf J}}=-F_{
1{\sf JJ}}j_{\sf J}(\rho) \left[Y_{{\sf
J}}(\hat{\rb})\otimes{\mbn}\right]_{{\sf J}}, \label{A3}\er with
${\sf L}\ge 0$, and ${\sf J}\ne 0$.
Thus \br {\sf O}_{\pm1{\sf J}} &=&i(-\gA \pm\gw) (\M^{\sss A,R}_{1\sf
J}\pm i{\M}^{\sss A,I}_{1{\sf J}})
 \nn\\
 &+&\gvs(\M^{\sss V,R}_{1\sf J}\pm i{\M}^{\sss  V,I}_{1{\sf J}}),
\label{A4}\er and writing \br {\sf O}_{\emptyset{\sf
J}}&=&{\sf O}_{\emptyset{\sf J}}^{\sss R}+i{\sf O}_{\emptyset{\sf J}}^{\sss I},
\nn\\
{\sf O}_{m{\sf J}}&=&{\sf O}_{m{\sf J}}^{\sss R}+i{\sf O}_{m{\sf J}}^{\sss I},
\label{A5}\er  it is not difficult to discover
that: \bit \item {\it For natural parity states }, with
$\pi=(-)^J$,  i.e., $J^\pi=0^+,1^-,2^+,3^-,\cdots$: \br
 {\sf O}_{\emptyset{\sf J}}^{\sss R}&=&g_{{\sss{V}}}\M_{\sf J}^{\sss V},
\nn\\
 {\sf O}_{{0}{\sf J}}^{\sss R}&=&\frac{{\tilde k}_{\emptyset}}{\k}\gV\M_{\sf J}^{\sss V},
\nn\\
{\sf O}_{\pm1{\sf J}}^{\sss R} &=&(\pm\gA -\gw){\M}^{\sss A,I}_{1{\sf
J}}+\gvs\M^{\sss V,R}_{1\sf J},
\label{A6}\er
\item {\it For unnatural parity states}, with $\pi=(-)^{J+1}$,  i.e.,
$J^\pi=0^-,1^+,2^-,3^+,\cdots$: \br {\sf O}_{\emptyset{\sf J}}^{\sss
I}&=&-\gas\M^{\sss A}_{\sf J} -(\ga+\gpa)\M^{\sss A}_{0{\sf J}},
\nn\\
{\sf O}_{{0}{\sf J}}^{\sss I} &=&(\gA -\gpb)\M^{\sss A}_{{0 \sf  J}},
\nn\\
{\sf O}_{\pm1{\sf J}}^{\sss I} &=&(\gA \mp\gw)\M^{\sss A,R}_{1\sf J}
\mp\gV{\M}^{\sss  V,I}_{1{\sf J}}.
 \label{A7}\er
  \eit
   These operators have to be used in \rf{2.22}, instead of those defined in
\rf{2.12}, and  \rf{2.20}.

The correspondence
 between the individual matrix elements, defined by Donnelly, and
Peccei in Eq (3.31)of Ref.~\cite{Don79}, and the ones used here,  is:
\br
 M_{\sf  J}&\go&\M^{\sss V}_{\sf  J},
\nn\\
\Delta_{\sf  J}&\go &\sqrt{2}{\M}^{\sss  V,I}_{1{\sf J}},
\nn\\
\Delta'_{\sf  J}&\go &-\sqrt{2}{\M}^{\sss  V,R}_{1{\sf J}},
\nn\\
\Sigma_{\sf  J}&\go &\sqrt{2}{\M}^{\sss  A,I}_{1{\sf J}},
\label{A8}\\
\Sigma'_{\sf  J}&\go &-\sqrt{2}{\M}^{\sss  A,R}_{1{\sf J}},
\nn\\
\Sigma_{\sf  J}"&\go &{\M}^{\sss  A}_{0{\sf J}},
\nn\\
\Omega_{\sf  J}&\go &{\M}^{\sss  A}_{{\sf J}}. \nn\er

Moreover, the correspondence between the linear combinations of
these matrix elements defined in~\cite[Eqs. (3.32)]{Don79}
 (for ${\hat L}_{\sf  J}$ see ~\cite[Eq. (14)]{Hax79}), and
the ones introduced here is:
  \bit
\item {\it For natural parity states }:
\br {\hat M}_{\sf  J}&=& {\sf O}_{\emptyset{\sf J}},
\nn\\
{\hat L}_{\sf  J} &=& {\sf O}_{0{\sf J}},
\nn\\
{\hat T}_{\sf  J}^{\rm el}\pm {\hat T}_{\sf  J}^{\rm mag5}
&=&-\sqrt{2}{\sf O}_{\pm1{\sf J}},
 \label{A9}\er
\item {\it For unnatural parity states}: \br {\hat M}_{\sf
J}^5&=&{\sf O}_{\emptyset{\sf J}},
\nn\\
-i{\hat L}_{\sf  J}^5&=&{\sf O}_{0{\sf J}},
\nn\\
i({\hat T}_{\sf  J}^{\rm el5}\pm {\hat T}_{\sf  J}^{\rm
mag})&=&\sqrt{2}{\sf O}_{\pm1{\sf J}}. \label{A10}\er \eit The
following relation can also be useful: \br
 {\sf O}_{\emptyset{\sf  J}}&=&{\hat \M}_{\sf  J},
\label{A11}\\
 {\sf O}_{m\sf J}&=& \left\{
\begin{array}{ccc}
{\hat \L}_{\sf  J},\;\;& \mbox{for} \;& m=0 \\
-{1\over \sqrt{2}}\left[m{ \hat{\T}^{\rm mag}}_{\sf  J} +{\hat
\T^{\rm el}}_{\sf  J}\right],\;\;& \mbox{for}\;& m=\pm 1
\end{array}\right.,
\nn\er where ${\hat \M}_{\sf  J}={\hat M}_{\sf  J}+{\hat M}_{\sf
J}^{5}$, ${\hat \L}_{\sf  J}={\hat L}_{\sf  J}+{\hat L}_{\sf
J}^{5}$, ${\hat \T^{\rm el}}_{\sf  J}={\hat T}_{\sf  J}^{\rm
el}+{\hat T}_{\sf  J}^{\rm el5}$, and ${\hat \T^{\rm mag}}_{\sf
J}={\hat T}_{\sf  J}^{\rm mag}+{\hat T}_{\sf  J}^{\rm mag5}$.

The matrix elements of Kuramoto \etal~\cite{Kur90} are related
with our non-relativistic operators \rf{2.14} as: \br |\bra{f}{\hat
1}\ket{i}|^2&=&4\pi \sum_{{\sf J}^\pi_n} |\Bra{{\sf J}^\pi_n}
\M^{\sss V}_{\sf J}\Ket{0^+}|^2,
\nn\\
|\bra{f}{\hat \sigma}\ket{i}|^2&=&4\pi \sum_{{\sf
J}^\pi_n}\sum_{m=0,\pm1} |\Bra{{\sf J}^\pi_n}\M^{\sss A}_{{m\sf
J}}\Ket{0^+}|^2,
\nn\\
\Lambda&=&\frac{4\pi}{3} \sum_{{\sf J}^\pi_n} \left[|\Bra{{\sf
J}^\pi_n}\M^{\sss A}_{{0\sf J}}\Ket{0^+}|^2\right.
\nn\\
&-&\left.|\Bra{{\sf J}^\pi_n}\M^{\sss A}_{{1\sf
J}}\Ket{0^+}|^2\right]. \label{A12}\er In Ref.~\cite{Kur90} are
neglected the relativistic operators $\M^{\sss A}_{\sf J}$, and
$\M^{\sss V}_{{m\sf J}}$ defined in \rf{2.15}.

\section{Sum Rule Approach  \label{B}}
We follow here
 the  sum-rule  approach developed by Kuramoto \etal~\cite{Kur90}, and adapt it to our formalism.
We  start from Eqs. \rf{2.25}, and \rf{2.27}, and as in this work we assume  that  the
$\w_{{\sf J}^\pi_n}$  dependence  of  the  integrand  can  be
ignored, fixing  it at  a  representative  value
$\overline{\w_{{\sf J}^\pi_n}}$ .  The  summation  over  final  nuclear
states  ${\sf J}^\pi_n$ then  can  be  carried  out  by  closure,
and the ICS is
\br
\s_\ell^{SR}(E_{\nu})& = &G^2\frac{|\pb_\ell|
E_\ell}{2\pi} F(Z+S,E_\ell)
\nn\\
&\x&\int_{-1}^1d(\cos\theta)\T^{SR},
\label{B1}\er
where the lepton energy is $E_\ell=E_\nu-\overline{\w_{{\sf J}^\pi_n}}$, while the sum-rule
matrix element reads:
\br
\T^{SR}&=& \sum_{\a=\emptyset,0,\pm 1}
 \bra{0^+}O_\a^\dag O_\a\ket{0^+}\L_{\a}
\nn\\
&-&2\Re\left( \bra{0^+}O_{\emptyset }^\dag O_{ 0}\ket{0^+}\L_{\emptyset 0}\right).
\label{B2}\er
The operators $O_\a$ are given by \rf{2.6}, and the lepton traces by
\cite[Eq. (2.24)]{Krm05}. The matrix elements in \rf{B2} are proportional to
$N(1-D)$, where $N_N$ is  the number of neutrons (protons), contained in the
target nucleus for the neutrino (anti-neutrino) reaction. The  correlation  functions
$D$ come from the Pauli-exclusion-effect, and depend on the type of the operator. One gets:
\br
\T^{SR}&=&N_N\left(T_{\emptyset}\L_{\emptyset}
+\sum_{M}T_M\L_{M}-2T_{\emptyset 0}\L_{\emptyset 0}\right),
 \label{B3}\er
with
\br
 T_\emptyset&\equiv&\gV^2(1-D_S) + (\ga+\gpa)^2(1-D_L),
 \nonumber\\
T_{0}&\equiv&\gv^2(1-D_S) +(\gA -\gpb)^2(1-D_L),
 \nonumber\\
T_{1}&\equiv&(\gA -\gw)^2(1-D_T),
\label{B4}\\
T_{-1}&\equiv&(\gA +\gw)^2(1-D_T),
 \nonumber\\
T_{\emptyset 0}&\equiv&-\gV\gv(1-D_S) + (\ga+\gpa)(\gA -\gpb)(1-D_L).
 \nonumber\er
The  correlation  functions $D_S, D_L$ and $D_S$ were taken from the
SM calculation done by  Bell, and Llewellyn Smith~\cite{Bel71} with HO  wave  functions, and representing
the   nuclear  ground  state  by  a single determinant  wave  function.
The results for $^{12}$C are ~\cite[Table 1]{Bel71}):
\br
D_S&=&e^{-\eta}\left[1+0.148\eta^2\right],
\nn\\
D_T&=&e^{-\eta}\left[0.704+0.148\eta+0.148\eta^2\right],
\nn\\
D_L&=&e^{-\eta}\left[0.704+0.296\eta+0.148\eta^2\right],
\label{B5}\er
where $\eta=\frac{1}{2} b^2\k^2\cong 0.0558$.

As seen from \rf{2.26},  the factor $|\pb_\ell|E_\ell$ in \rf{B1}
behaves as $(E_\nu-\overline{\w_{{\sf J}^\pi_n}})^2$,
and therefore $\s_\ell^{SR}(E_{\nu})$ depends very critically on the
average value for the excitation energy $\overline{\w_{{\sf J}^\pi_n}}$.
\br
E_\ell&=&E_\nu-\w_{{\sf J}^\pi_n},~
|\pb_\ell|=\sqrt{(E_{\nu}-\w_{{\sf J}^\pi_n})^2-m_\ell^2},
\nn\\
\kappa&=&|\pb_\ell-\qb_{\nu}|\nn\\
&=&\sqrt{2E_{\nu}(E_\ell-|\pb_\ell|\cos\theta)-m_\ell^2
+\w_{{\sf J}^\pi_n}^2},
\label{211}\er

\section{Extreme Relativistic Limit \label{C}}
 Using the present formalism the ERL, defined by the limit of the lepton
 velocity $|\pb_\ell|/E_\ell\go 1$,
yields \br \s^{ERL}_\ell(E_{\nu}) &=& \sum_{J^{\pi}_n}\frac{
E^2_\ell}{2\pi} F(Z+S,E_\ell) \int_{-1}^1
d(\cos\theta)\T^{ERL}_{{\sf J}^\pi_n}(\kappa), \nn\\\label{C1}\er
with \br \kappa&=&\sqrt{2E_{\nu}E_\ell(1-\cos\theta)+\w_{{\sf
J}^\pi_n}^2}, \label{C2}\er and \br \T^{ERL}_{{\sf
J}^\pi_n}(\k)&=&{4\pi G^2} \left[
2\cos^2\frac{\theta}{2}\left|\Bra{{\sf J}^\pi_n}{\sf
O}_{\emptyset{{\sf J}}}(\k)
 -\frac{k_{\emptyset}}{\k }{\sf O}_{0{\sf J}}(\k)\Ket{0^+}\right|^2 \right.
\nn\\
&+&\sum_{m=\pm 1}|\Bra{{\sf J}^\pi_n}{\sf O}_{m{\sf
J}}(\k)\Ket{0^+}|^2
\nn\\
&\x&\left(\frac{k^2}{\k^2} \cos^2\frac{\theta}{2}+2\sin^2
\frac{\theta}{2}\right.
\nn\\
&+&2m S\left.\left. \sin\frac{\theta}{2}\sqrt{\frac{k^2}{\k^2}
\cos^2\frac{\theta}{2}+\sin^2\frac{\theta}{2}}\right)\right].
\label{C3}\er

\section{Muon Capture rate \label{D}}
For the sake of completeness we also show the formula
for the muon capture process within the present formalism.
 Here $\k=E_\nu=\mass_\mu-\w_{{\sf J}^\pi_n}-\Delta\Mass-E_B$, where
  $E_B^\mu$ is the
binding energy of the muon in the $1S$ orbit, and instead of
\rf{2.5} one has: \br
\gv&=&\gV\frac{E_\nu}{2\Mass};~\ga=\gA\frac{E_\nu}{2\Mass},\nn\\
\gw&=&(\gV+\gM)\frac{E_\nu}{2\Mass};~ \gp=\gP\frac{E_\nu}{2\Mass},
\label{D1}\er where $\gp=\gpb-\gpa$. The muon capture transition
rate reads \br \Lambda(\w_{{\sf
J}^\pi_n})&=&\frac{E_\nu^2}{2\pi}|\phi_{1S}|^2
\T_{\Lambda}(\w_{{\sf J}^\pi_n}), \label{D2}\er where $\phi_{1S}$
is the  muonic bound state wave function evaluated at the origin,
and \br \T_{\Lambda}(\w_{{\sf J}^\pi_n})&=&4\pi G^2
\left[|\Bra{{\sf J}^\pi_n}{\sf O}_{\emptyset{ {\sf J}}}(E_\nu)-
{\sf O}_{0{ {\sf J}}}(E_\nu)\Ket{0^+}|^2\right.
 \nn\\
 &+&\left.2|\Bra{{\sf J}^\pi_n}{\sf O}_{-1{ {\sf J}}}(E_\nu)
 \Ket{0^+}|^2\right],
\label{D3}\er with \bit \item {\it For natural parity states },
with $\pi=(-)^J$,  i.e., $J^\pi=0^+,1^-,2^+,3^-,\cdots$: \br {\sf
O}_{\emptyset{\sf J}}-{\sf O}_{0, \sf  J}&=&
\gV\frac{\mass_\mu-\Delta E_{\rm Coul}-E_B}{E_\nu}\M^{\sss V}_{\sf
J},
\nn\\
{\sf O}_{-1{\sf J}} &=&-(\gA +\gw){\M}^{\sss A,I}_{-1{\sf J}}
+\gvs\M^{\sss V,R}_{-1\sf J}, \label{D4}\er
and
 \item {\it For
unnatural parity states}, with $\pi=(-)^{J+1}$,  i.e.,
$J^\pi=0^-,1^+,2^-,3^+,\cdots$: \br {\sf O}_{\emptyset{\sf
J}}-{\sf O}_{0, \sf  J}&=&\gas\M^{\sss A}_{\sf J}+ \left(\gA
+\ga-\gp\right)\M^{\sss A}_{0{\sf J}},
\nn\\
{\sf O}_{-1{\sf J}} &=&-(\gA +\gw)\M^{\sss A,R}_{-1\sf J}
-\gvs{\M}^{\sss  V,I}_{-1{\sf J}}. \label{D5}\er \eit

\end{appendix}


\begin{thebibliography}{99}
\bibitem{Ath96} C.~Athanassopoulos \etal~ [LSND Collaboration],
                Phys. Rev. C {\bf 54}, 2685 (1996); {\it ibid}
                Phys. Rev. Lett. {\bf 77}, 3082 (1996).

\bibitem{Ath98} C.~Athanassopoulus \etal~ [LSND Collaboration],
                Phys. Rev. C {\bf 58}, 2489 (1998); {\it ibid}
                Phys. Rev. Lett. {\bf 81}, 1774 (1998).
\bibitem{Agu01} A.~Aguilar \etal~ [LSND collaboration],
                Phys. Rev. D {\bf 64}, 112007 (2001).
\bibitem{Fuk98} Y. Fukuda \etal~ [Super-Kamiokande  Collaboration],
                Phys. Rev. Lett. {\bf 81}, 1562 (1998);
                Y. Ashie  \etal~ [Super-Kamiokande Collaboration],
                Phys. Rev. Lett. {\bf 93}, 101801 (2004).
\bibitem{Aha05} B. Aharmim \etal~ [SNO Collaboration],
                Phys. Rev. C {\bf 59}, 055502 (2005);
                M. B. Smy \etal~ [Super-Kamiokande Collaboration],
                Phys. Rev. D {\bf 69}, 011104 (2004).
\bibitem{Ara04} T. Araki {\it et al.} [KamLAND Collaboration],
                Phys. Rev. Lett. {\bf 94}, 081801 (2005).
\bibitem{Ahn03} M. H. Ahn \etal ~ [K2K Collaboration],
                Phys. Rev. Lett. {\bf 90}, 041801 (2003).
\bibitem{Mas98} R. Maschuw \etal~  [KARMEN Collaboration],
                Prog. Part. Phys. {\bf 40}, (1998) 183; and references
                therein mentioned.
\bibitem{Arm02} B. Armbruster \etal ~ [KARMEN collaboration],
                Phys. Rev. D {\bf 65}, 112001 (2002).
\bibitem{All90} R. C. Allen \etal~,
                Phys. Rev. Lett. {\bf 64}, 1871 (1990).
\bibitem{Kra92} D. A. Krakauer \etal~,
                Phys. Rev. C {\bf 45}, 2450 (1992).
\bibitem{SciBar}A. A. Aguilar-Arevalo \etal (SciBooNE Collaboration),
arXiv:hep-ex/0601022.
\bibitem{MiniBooNE} A. A. Aguilar-Arevalo et al. (MiniBooNE Collaboration), Nucl.
Instrum. Methods A {\bf 599}, 28 (2009).
\bibitem{Efr05} Y. Efremenko,
                Nucl. Phys. {\bf B138}(Proc. Suppl.), 343 (2005);
                F.T. Avignone III and Y.V. Efremenko,
                J. Phys. G {\bf 29}, 2615 (2003).
\bibitem{Aga07} N.Yu. Agafonova \etal,
                Astron. Phys. {\bf 27}, 254 (2007).
\bibitem{Fuk88} M. Fukugita, Y. Kohyama and K. Kubodera,
                Phys.  Lett. {\bf B212}, 139 (1988).
\bibitem{Aut07} D. Autiero \etal,
                J. Cosmol. Astropart. Phys. \textbf{0711}, 011 (2007); arXiv:hep-ph/0705.0116.
\bibitem{Lun03} C. Lunardini and A. Y. Smirnov,
     J. Cosmol. Astropart. Phys. \textbf{0306}, 009 (2003).
\bibitem{Dig03} A. S. Dighe, M. T. Keil, and G. G. Raffelt,
     J. Cosmol. Astropart. Phys. \textbf{0306}, 005 (2003).
\bibitem{Dua07} H. Duan, G. M. Fuller, J. Carlson, and Y.-Q. Zhong,
     Phys. Rev. Lett. \textbf{99}, 241802 (2007).
\bibitem{Das08} B. Dasgupta, A. Dighe, and A. Mirizzi,
     Phys. Rev. Lett. \textbf{101}, 171801 (2008).
\bibitem{Das10} B. Dasgupta, A. Mirizzi, I. Tamborra, and R. Tomas,
    Phys. Rev.  \textbf{D81}, 093008 (2010).
\bibitem{Mez10} M. Mezzetto and T. Schwetz,
        arXiv:1003.5800v1 [hep-ph] (2010).
\bibitem{Str06} A. Strumia and F. Vissani, arXiv: hep-ph/0606054v2.
\bibitem{Ath97} C.~Athanassopoulus {\it et al.} [LSND Collaboration],
               Phys. Rev. C  {\bf 55}, 2078 (1997).
\bibitem{Aue01} L. B. Auerbach \etal [LSND Collaboration],
                Phys. Rev. C {\bf 64}, 065501 (2001).
\bibitem{Zei98} B. Zeitnitz \etal [KARMEN Collaboration],
                Prog. Part. Nucl. Phys. \textbf{40}, 169 (1998).
\bibitem{Ath97a} C.~Athanassopoulus {\it et al.} [LSND Collaboration],
                 Phys. Rev. C {\bf 56}, 2806 (1997).
\bibitem{Aue02a} L. B. Auerbach \etal [LSND Collaboration],
                 Phys. Rev. {\bf C 66}, 015501 (2002).
\bibitem{LSND} LSND home page,
               http://www.nu.to.infn.it/exp/all/lsnd/

\bibitem{MiniBooNE Collaboration} Teppei Katori, in The
5th International Workshop on Neutrino-Nucleus
Interactions in the Few-GeV Region, edited by Geralyn
P. Zeller, Jorge G. Morfin, Flavio Cavanna, AIP Conf.
Proc. No. 967 (AIP, New York, 2007), p. 123;
A.A. Aguilar-Arevalo \etal, Phys. Rev. Lett. {\bf 103}, 081801 (2009);
 Phys. Rev. D {\bf 81}, 013005 (2010).
\bibitem{K2K Collaboration} A. Rodriguez, \etal,  Phys. Rev. D {\bf 78}, 032003 (2008).
\bibitem{SciBooNE Collaboration}Y. Kurimoto~\etal,  Phys. Rev. D {\bf 81}, 033004 (2010).

\bibitem{Con72} J.S. O'Connell, T.W. Donelly and
                J.D. Walecka, Phys. Rev. C {\bf 6}, 719 (1972).
\bibitem{Don70} T.W. Donelly, Phys. Rev. C {\bf 1}, 853 (1970).
\bibitem{Bro85} B.A. Brown and B.H. Wildenthal,
                At. Data Nucl. Data Tables {\bf 33}, 347 (1985).
\bibitem{Cas87} H. Castillo and F. Krmpoti\'c,
                Nucl. Phys. {\bf A469}, 637 (1987).
\bibitem{Ost92} F. Osterfeld, Rev. Mod. Phys. {\bf 64}, 491 (1992), and references therein.
\bibitem{Mar96} G. Mart\'inez-Pinedo \etal,
                Phys. Rev. C \textbf{53}, R2602 (1996).
\bibitem{Kol94} E. Kolbe, K. Langanke and S. Krewald,
                Phys. Rev. C {\bf 49}, 1122 (1994).
\bibitem{Kol94a} E. Kolbe, K. Langanke and P.Vogel,
                 Phys. Rev. C {\bf 50}, 2576 (1994).
\bibitem{Hay00} A.C. Hayes and I.S. Towner,
                Phys. Rev. C {\bf 61}, 044603 (2000).
\bibitem{Vol00} C. Volpe, N. Auerbach, G. Col\`o, T. Suzuki,
                N. Van Giai, Phys. Rev. C {\bf 62}, 015501 (2000).
\bibitem{Suz06} T. Suzuki \etal,
                Phys. Rev. C {\bf 74},034307 (2006).
\bibitem{Mil72} G. H. Miller \etal ,
                Phys. Lett. {\bf B41}, 50 (1972).
\bibitem{Mea01} D.F. Measday, Phys. Rep. {\bf 354}, 243 (2001).
\bibitem{Sto02} T.J. Stocki \etal, Nucl. Phys. {\bf A697}, 55 (2002).
\bibitem{Krm02} F. Krmpoti\'c, A. Mariano and A. Samana,
                Phys.Lett. {\bf B541}, 298 (2002).
\bibitem{Krm05} F. Krmpoti\'c, A. Samana, and A. Mariano,
                Phys. Rev. C {\bf 71}, 044319 (2005).
\bibitem{Sam06} A. Samana, F. Krmpoti\'c, A. Mariano and R. Zukanovich
                Funchal, Phys. Lett. {\bf B642}, 100 (2006).
\bibitem{Paa07} N. Paar, D. Vretenar, T. Marketin and P. Ring,
                Phys. Rev. C {\bf 77}, 024608 (2008).
\bibitem{Mar09} T. Marketin, N. Paar, T. Nik\v si\'c and D. Vretenar,
                Phys. Rev. C {\bf 79}, 054323 (2009).
\bibitem{Hag01} K. Hagino and  H. Sagawa,
                Nucl.Phys. {\bf A695}, 82 (2001).
\bibitem{Rod08}V. Rodin and A. Faessler, Phys.Rev. C {\bf 77}, 025502
(2008).

\bibitem{Smi72} R.A.  Smith and  E.J. Moniz,
                Nucl. Phys. {\bf B43}, 605  (1972).
\bibitem{Nie04} J. Nieves, J.E. Amaro, and M. Valverde,
                Phys. Rev. C {\bf 70}, 055503 (2004).
\bibitem{Val06} M. Valverde, J.E. Amaro, and J. Nieves,
                Phys. Lett. {\bf B638}, 325 (2006).
                Phys. Rev. C {\bf 77}, 025502 (2008).
\bibitem{Ma85} C. Mahaux, P.E. Bortignon, R.A. Broglia, and C.H.
               Dasso, Phys. Rep. {\bf 120}, 1 (1985).
\bibitem{Ja73} G. Jacob and T. A. J. Maris,
               Rev. Mod. Phys. {\bf 45}, 6 (1973).
\bibitem{Fr84} S. Frullani and J. Mougey,
               Adv. Nucl. Phys. \textbf{14}, 1 (1984).
\bibitem{Be85} S.L. Belostotskii \etal, Sov. J.
               Nucl. Phys. \textbf{41}, 903 (1985);
               S.S. Volkov \etal, Sov. J.
               Nucl. Phys. \textbf{49}, 848 (1990).
\bibitem{Le94} M. Leuschner \etal,
               Phys. Rev. C  \textbf{49}, 955 (1994).
\bibitem{Ya96} T. Yamada, M. Takahashi, and K. Ikeda,
               Phys. Rev. C \textbf{53}, 752 (1996).
\bibitem{Ya01} T. Yamada, Nucl. Phys. \textbf{A687}, 297c (2001).
\bibitem{Yo03} M. Yosoi \etal,
               Phys. Lett. \textbf{B551}, 255 (2003).
\bibitem{Ya04} T. Yamada, M. Yosoi, and H. Toyokawa,
               Nucl. Phys. \textbf{A738}, 323  (2004).
\bibitem{Ko06} K. Kobayashi \etal,
               arXiv:nucl-ex/0604006.

\bibitem{Ama05} J. E. Amaro, M. B. Barbaro, J. A. Caballero, T. W. Donnelly,
and C. Maieron,  Phys. Rev. C {\bf 71}, 065501 (2005).
\bibitem{Kol03} E. Kolbe, K. Langanke, G. Mart\'inez-Pinedo and  P. Vogel,
 J. Phys. {\bf G29}, 2569 (2003).
\bibitem{Kim08} K. S. Kim, Myung-Ki Cheoun, and Byung Geel Yu,
Phys. Rev. C {\bf 77}, 054604 (2008).
\bibitem{But10} A. V. Butkevich, Phys. Rev. C {\bf 78}, 015501 (2008);
 Phys. Rev. C {\bf 80}, 014610 {2009}; arXiv:1006.1595.
 \bibitem{Lei09} T. Leitner, O. Buss, L. Alvarez-Ruso, and U. Mosel
Phys. Rev. C {\bf 79}, 034601 (2009).
\bibitem{Mart09} M. Martini, M. Ericson, G. Chanfray and J. Marteau,
                 Phys. Rev. C \textbf{80}, 065501 (2009).
\bibitem{Kur90} T. Kuramoto \etal, Nucl. Phys.  {\bf A512}, 711 (1990).

\bibitem{Vog86} P. Vogel and M.R. Zirnbauer,
                Phys. Rev. Lett. {\bf 57}, 3148 (1986).
\bibitem{Cha87} D. Cha,
                Phys. Rev. {\bf C27}, 2269  (1983).

\bibitem{Don79} T. W. Donnelly and R. D. Peccei,
                Phys. Rep. {\bf 50}, 1 (1979).
\bibitem{Wal95} J.D. Walecka, {\it Theoretical Nuclear and Subnuclear
                Physics}, {\it Oxford University Press, New York},
                531 (1995).
\bibitem{Hax79} T.W. Donnelly and W.C. Haxton,
                Atomic Data and Nuclear Data Tables {\bf 23},
                103 (1979).


\bibitem{Krm80} F. Krmpoti\'c, K. Ebert, and W. Wild,
Nucl.Phys. {\bf A342}, 497 (1980);  F. Krmpoti\'c, Phys. Rev. Lett.
{\bf 46}, 1261  (1981).

\bibitem{Beh82} H. Behrens and W. B\"{u}hring,
              {\it Electron Radial Wave Functions
              and Nuclear Beta Decay} (Clarendon, Oxford, 1982), and references therein.
\bibitem{Bli66} R.J. Blin-Stoyle and S.C.K. Nair,
                Advances in Physics {\bf 15},  493 (1966).
\bibitem{Eng98} J. Engel, Phys. Rev. C {\bf 57}, 2004 (1998).
\bibitem{Sam08} A.R. Samana  and C.A. Bertulani,
                Phys. Rev. C {\bf 78}, 024312 (2008)
\bibitem{Hir90a} J. Hirsch and F. Krmpoti\'c,
                 Phys. Rev. C {\bf 41}, 792 (1990), {\it ibid}
                 Phys. Lett. {\bf B246}, 5 (1990).
\bibitem{Krm92} F. Krmpoti\'c, J. Hirsch and H. Dias,
                Nucl. Phys. {\bf A542}, 85 (1992).
\bibitem{Krm93} F. Krmpoti\'c, A. Mariano, T.T.S. Kuo, and
                K. Nakayama,  Phys. Lett. {\bf B319}, 393 (1993).
\bibitem{Krm94} F. Krmpoti\'c and Shelly Sharma,
                Nucl. Phys. {\bf A572}, 329 (1994).
\bibitem{Sam09} A.R. Samana, F. Krmpoti\'c  and C.A. Bertulani,
                Comp. Phys. Comm. \textbf{181}, 1123 (2010).
\bibitem{PNVR.04} N. Paar, T. Nik{\v{s}}i{\'{c}}, D. Vretenar,
                  and P. Ring, Phys. Rev. C \textbf{69}, 054303 (2004).
\bibitem{LNVR.05} G. A. Lalazissis, T. Nik\v si\' c, D. Vretenar,
                  and P. Ring, Phys. Rev. C \textbf{71}, 024312 (2005).
\bibitem{BGG.91} J. F. Berger, M. Girod, and D. Gogny,
                 Comp. Phys. Comm.\textbf{ 63}, 365 (1991).
\bibitem{Paa.03} N. Paar, P. Ring, T. Nik\v si\' c, and D. Vretenar,
                 Phys. Rev. C \textbf{67}, 034312 (2003).
\bibitem{PVKC.07} N. Paar, D. Vretenar, E. Khan, and G. Col\`o,
                  Rep. Prog. Phys. \textbf{70}, 691 (2007).
\bibitem{Ajz85} F. Ajzenberg-Selove, Nucl. Phys. {\bf A 433}, 1(1985);
                TUNL Nuclear Data Evaluation Project.
                Webpage: http:// www.tunl.duke.edu/nucldata/.
\bibitem{Al78} D. E. Alburger and A.M. Nathan,
               Phys. Rev. C {\bf 17}, 280 (1978).
\bibitem{Str09} A. Strumia and F. Vissani,
                arXiv.org/abs/hep-ph/0606054
\bibitem{Eng96} J. Engel, E. Kolbe, K. Langanke,and  P. Vogel,
                Phys. Rev. C {\bf 54}, 2740 (1996).
\bibitem{Kol99a} E. Kolbe, K. Langanke and G. Mart\'{\i}nez-Pinedo,
                 Phys. Rev. C {\bf 60}, 052801(R) (1999).
\bibitem{Kr83}  F.  Krmpoti\'c, K. Nakayama, and A. P. Gale\~ao,
               Nucl. Phys. {\bf A339}, 475 (1983).
\bibitem{Bud03} H. Budd, A. Bodek, and  J. Arrington, arXiv:hep-ex/0308005.

\bibitem{Kol99b} E. Kolbe, K. Langanke and P. Vogel
                 Nucl. Phys. {\bf A652}, 91 (1999).

\bibitem{Mart10} M. Martini, M. Ericson, G. Chanfray and J. Marteau,
                 Phys. Rev. C \textbf{81}, 045502 (2010).

\bibitem{Bau10} E. Bauer, and  G. Garbarino, Phys.Rev. C \textbf{81}, 064315 (2010).
\bibitem{Ma91} A. Mariano, E. Bauer, F. Krmpoti\'c, and A.F.R. de Toledo Piza,
                Phys.Lett. {\bf B268}, 332 (1991).
\bibitem{Va92} D. Van Neck, M. Waroquier, V. Van der Sluys, and J. Ryckebusch, Phys. Lett.
 {\bf B274}, 142(1992).
\bibitem{Ma95} A. Mariano,  F. Krmpoti\'c, and A.F.R. de Toledo Piza,
               Phys. Rev. C \textbf{49}, 2824 (1994), and  Phys. Rev. C \textbf{53}, 1664 (1996).

\bibitem{Laz07} R. Lazauskas and C. Volpe,
                Nucl. Phys. {\bf A792}, 219 (2007).

\bibitem{Vol04} C. Volpe,
               J. Phys. G \textbf{30}, L1 (2004); arXiv:hep-ph/0303222.
\bibitem{Woo90} S. E. Woosley, D. H. Hartmann, R. D. Hoffman and
                W. C. Haxton, Ap. J.{\bf 356}, 272 (1990).
\bibitem{Kei03} M. Th. Keil, G. G. Raffelt, and
                H. -Th. Janka, Ap. J.{\bf 590}, 971 (2003).
\bibitem{Akh00} E.Kh. Akhmedov, Lectures given at Trieste
                Summer School in Particle Physics, June 7-9, 1999;
                arXiv:hep-ph/0001264v2.
\bibitem{Dun07} H. Duan, G. M. Fuller, J. Carlson, and Y-Z. Qian,
                Phys. Rev. Lett. {\bf 99}, 241802 (2007); H. Duan,
                G. M. Fuller and Y.Z. Qian, J. Phys. G \textbf{36}, 113201 (2009).
\bibitem{Na82} K. Nakayama,  A. P. Gale\~ao, and F.  Krmpoti\'c,
 Phys.  Lett. {\bf B114}, 217 (1982).
\bibitem{Ch10} M. K Cheoun~\etal, Phys. Rev. C {\bf 81}, 028501 (2010).
\bibitem{Bel71} J.S. Bell and C.H. Llewellyn Smith,
                Nucl. Phys. {\bf B28}, 317 (1971).
%
\end{thebibliography}
\end{document}